\documentclass[a4paper,11pt]{article}

\usepackage{jcappub} 

\usepackage{epsfig}
\usepackage{amsmath}
\usepackage{amssymb}
\usepackage{natbib}
\usepackage{subfigure}
\usepackage{longtable}
\usepackage{graphicx}
\usepackage{textcomp}
\usepackage{color}
\usepackage{url}
\usepackage{relsize}

\title{\boldmath Extragalactic sources in Cosmic Microwave Background maps}

   \author[a,b,1]{G. De Zotti,\note{Corresponding author.}}
   \author[a]{G. Castex,}
   \author[c]{J. Gonz\'alez-Nuevo,}
   \author[c]{M. Lopez-Caniego,}
   \author[b]{M. Negrello,}
   \author[d]{Z.-Y. Cai,}
   \author[b]{M. Clemens,}
   \author[e]{J. Delabrouille,}
   \author[c]{D. Herranz,}
   \author[c]{L. Bonavera,}
   \author[f]{J.-B. Melin,}
   \author[g]{M. Tucci,}
   \author[h]{S. Serjeant,}
   \author[i,l]{M. Bilicki,}
   \author[m]{P. Andreani,}
   \author[n]{D. L. Clements,}
   \author[o]{L. Toffolatti}
   \author[p]{and B.F. Roukema}

\affiliation[a]{SISSA, Via Bonomea 265, 34136, Trieste, Italy}
\affiliation[b]{INAF-Osservatorio Astronomico di Padova, Vicolo dell'Osservatorio 5, I-35122 Padova, Italy}
\affiliation[c]{Instituto de F\'isica de Cantabria (CSIC-UC), Avda. los Castros s/n, 39005 Santander, Spain}
\affiliation[d]{Center for Astrophysics, University of Science and Technology of China, Hefei, 230026, China}
\affiliation[e]{APC, 10, rue Alice Domon et L\'eonie Duquet, 75205 Paris Cedex 13, France}
\affiliation[f]{DSM/Irfu/SPP, CEA-Saclay, F-91191 Gif-sur-Yvette Cedex, France}
\affiliation[g]{D\'epartement de Physique Th\'eorique and Center for Astroparticle Physics, Universit\'e de Gen\`eve, 24 quai Ansermet, CH–1211 Gen\`eve 4, Switzerland}
\affiliation[h]{Department of Physical Sciences, The Open University, Walton Hall, Milton Keynes MK7 6AA, UK}
\affiliation[i]{Astrophysics, Cosmology and Gravity Centre, Department of Astronomy, University of Cape Town, Private Bag X3, Rondebosch, South Africa}
\affiliation[l]{Kepler Institute of Astronomy, University of Zielona G\'{o}ra, ul. prof. Z. Szafrana 2, 65-246 Zielona G\'{o}ra, Poland}
\affiliation[m]{European Southern Observatory, Karl-Schwarzschild-Stra{\ss}e 2, D-85748, Garching, Germany}
\affiliation[m]{Astrophysics Group, Imperial College, Blackett Laboratory, Prince Consort Road, London SW7 2AZ, UK}
\affiliation[o]{Departamento de F\'isica, Universidad de Oviedo, C. Calvo Sotelo s/n, 33007 Oviedo, Spain
}
\affiliation[p]{Toru\'n Centre for Astronomy, Faculty of Physics, Astronomy and Informatics, Nicolaus Copernicus University, ul. Gagarina 11, 87-100 Toru\'n, Poland}

\emailAdd{gianfranco.dezotti@oapd.inaf.it}
\emailAdd{gcastex@sissa.it}
\emailAdd{gnuevo@ifca.unican.es}
\emailAdd{caniego@ifca.unican.es}
\emailAdd{mattia.negrello@oapd.inaf.it}
\emailAdd{zcai.cn@gmail.com}
\emailAdd{marcel.clemens@oapd.inaf.it}
\emailAdd{delabrouille@apc.univ-paris7.fr}
\emailAdd{herranz@ifca.unican.es}
\emailAdd{bonavera@ifca.unican.es}
\emailAdd{jean-baptiste.melin@cea.fr}
\emailAdd{Marco.Tucci@unige.ch}
\emailAdd{S.Serjeant@open.ac.uk}
\emailAdd{maciek@ast.uct.ac.za}
\emailAdd{pandrean@eso.org}
\emailAdd{d.clements@imperial.ac.uk}
\emailAdd{ltoffolatti@uniovi.es}

\abstract {We discuss the potential of a next generation space-borne CMB experiment for studies of extragalactic sources with reference to COrE$+$, a project submitted to ESA in response to the call for a Medium-size mission (M4). We consider three possible options for the telescope size: 1\,m, 1.5\,m and 2\,m (although the last option is probably impractical, given the M4 boundary conditions). The proposed instrument will be far more sensitive than \textit{Planck} and will have a diffraction-limited angular resolution. These properties imply that even the 1\,m telescope option will perform substantially better than \textit{Planck} for studies of extragalactic sources. The source detection limits as a function of frequency have been estimated by means of realistic simulations taking into account all the relevant foregrounds. Predictions for the various classes of extragalactic sources are based on up-to-date models. The most significant improvements over \textit{Planck} results are presented for each option. COrE$+$ will provide much larger samples of truly local star-forming galaxies (by about a factor of 8 for the 1\,m telescope, of 17 for 1.5\,m, of 30 for 2\,m), making possible analyses of the properties of galaxies (luminosity functions, dust mass functions, star formation rate functions, dust temperature distributions, etc.) across the Hubble sequence. Even more interestingly, COrE$+$ will detect, at $|b|> 30^\circ$, thousands of strongly gravitationally lensed galaxies (about 2,000, 6,000 and 13,000 for the 1\,m, 1.5\,m and 2\,m options, respectively). Such large samples are of extraordinary astrophysical and cosmological value in many fields. Moreover, COrE$+$ high frequency maps will be optimally suited to pick up proto-clusters of dusty galaxies, i.e. to investigate the evolution of large scale structure at larger redshifts than can be reached by other means. Thanks to its high sensitivity COrE$+$ will also yield a spectacular advance in the blind detection of extragalactic sources in polarization: we expect that it will detect up to a factor of 40 (1\,m option) or of 160 (1.5\,m option) more radio sources than can be detected by \textit{Planck} and, for the first time, from several tens (1\,m option) to a few hundreds  (1.5\,m option) of star forming galaxies. This will open a new window for studies of the global properties of magnetic fields in star forming galaxies and of their relationships with SFRs.   }


\notoc

\begin{document}
\maketitle
\flushbottom

\def\simlt{\mathrel{\rlap{\lower 3pt\hbox{$\sim$}}\raise 2.0pt\hbox{$<$}}}
\def\simgt{\mathrel{\rlap{\lower 3pt\hbox{$\sim$}} \raise
2.0pt\hbox{$>$}}}
\def\lsim{\,\lower2truept\hbox{${<\atop\hbox{\raise4truept\hbox{$\sim$}}}$}\,}
\def\gsim{\,\lower2truept\hbox{${> \atop\hbox{\raise4truept\hbox{$\sim$}}}$}\,}

\newcommand{\angstrom}{{\rm \mathring A}}



\section{Introduction}\label{par:intro}

We investigate the impact on studies of extragalactic sources of the planned fourth generation Cosmic Microwave Background (CMB) mission named Cosmic Origins Explorer plus (COrE$+$). Various options are being considered. The COrE$+$ ``light'' concept envisages an effective telescope size of $\simeq 1\,$m and a 60--600\,GHz frequency range with a total of 2040 detectors. More ambitious options, requiring substantive contributions from international partners (``COrE$+$ extended''), contemplate telescope sizes from $\simeq 1.5\,$m (baseline) to 2\,m (although the last option is unlikely to fit within the M4 constraints) and a frequency range from 60 to 800\,GHz (but we consider also the effect of an extension to 1200 GHz) with a total of 5800 detectors. The instrument  comprises more than two times as many frequency channels as \textit{Planck}, to make possible an efficient separation of foreground emissions. The very large number of state-of-the-art detectors ensures a much better sensitivity per channel.

The plan of the paper is the following.  In Section~\ref{sec:det_lim} we discuss the source detection in CMB maps. In Section~\ref{sect:general} we briefly describe the populations of extragalactic sources in the COrE$+$ frequency range. Section~\ref{sect:clumps} is about the extraction of the rich information content on sources below the detection limit. In Section~\ref{sect:polarization} we deal with counts in polarization. Finally, in Sect~\ref{sect:conclusions} we summarize our main results.

Throughout this paper we adopt the fiducial $\Lambda$CDM cosmology with best-fit \emph{Planck} $+$ WP $+$ high-resolution CMB data of the Atacama Cosmology Telescope (ACT) and South Pole Telescope (SPT) experiments as provided by
\cite{planck_parameters:2013}.

\
\begin{figure}
\includegraphics[width=0.48\textwidth, angle=0]{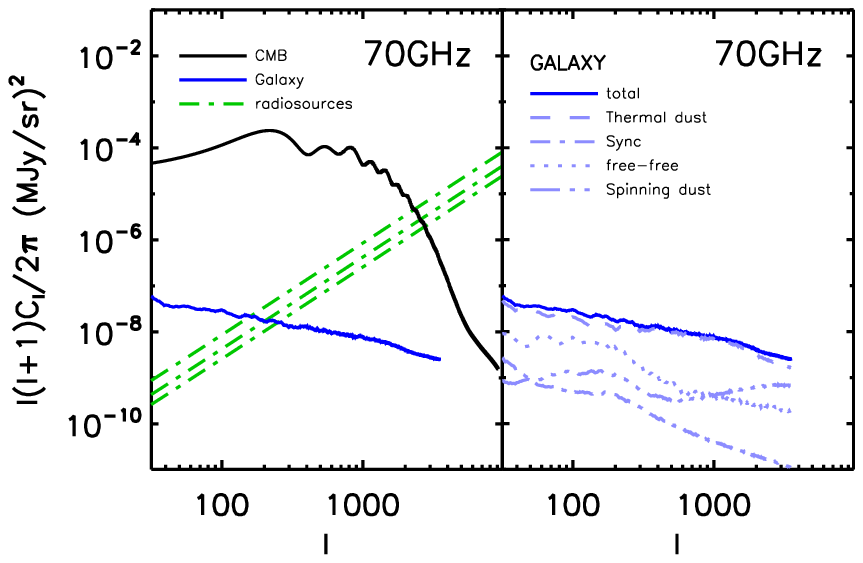}
\includegraphics[width=0.48\textwidth, angle=0]{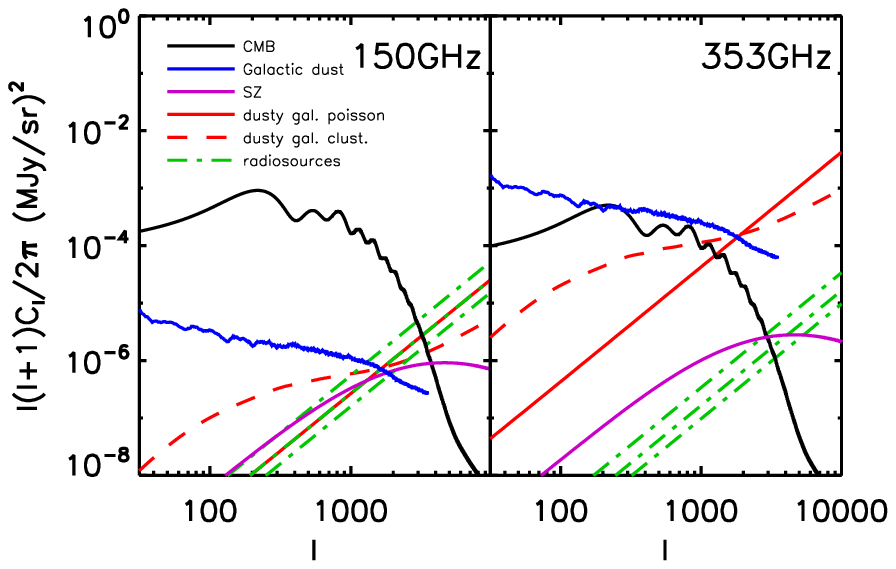}
\caption{\textbf{Left panel:} power spectra of the astrophysical components present in COrE$+$ maps at 70 GHz at high Galactic latitudes ($|b|\ge 30^\circ$), computed from the Planck Sky Model \citep{Delabrouille2013}, masking the region at $|b|<30^\circ$ with a $10^\circ$  apodization for $30^\circ< |b| < 40^\circ$. The three dot-dashed straight green lines show the power spectra of radio sources, that are the dominant extragalactic population at this frequency, computed after having masked sources brighter than the detection limits corresponding to the three options for the COrE$+$ telescope, shown in Fig.~\ref{fig:Slim_vs_sigma}. The contributions to the power spectrum of Galactic emission are detailed in the adjacent panel. \textbf{Right panel:} power spectra of the astrophysical components present in COrE$+$ maps at 150 and 353 GHz. The lines have the same meaning as in the left panel. However at these frequencies the only relevant Galactic emission is the one of thermal dust. Also, fluctuations from dusty galaxies, which have a large contribution from clustering, are increasingly important with increasing frequency.}
 \label{fig:power_spectrum}
\end{figure}
%

%

\section{Detection limits for CMB maps}\label{sec:det_lim}

\subsection{Basics}\label{sec:basics}

The development of algorithms for point source detection has a long history in astronomy. Images produced by CMB experiments have from this point of view important peculiarities \citep{HerranzVielva2010} that make inadequate the algorithms consecrated by decades of successful usage in many wavebands, from radio to X-rays, such as CLEAN \citep{Hogbom1974}, DAOFIND \citep{Stetson1992} and SExtractor \citep{BertinArnouts1996}. The peculiarities mostly follow from the limited angular resolution of CMB experiments.

In high resolution images intensity peaks on the angular scale of the Full Width at Half-Maximum (FWHM) of the telescope are due to either point sources or to detector noise and the highest significance peaks are straightforwardly interpreted as point sources. On the contrary, in CMB images also structure in diffuse emissions (the CMB itself and Galactic radiations) shows up on the FWHM scale, complicating the detection of point sources.

Optimal methods for source detection in CMB maps need the knowledge of the probability density functions (PDFs) of each signal contributing to the ``background'' \citep[for a review see][]{HerranzVielva2010}. In general, however, our knowledge of the PDFs is highly incomplete. A further complication comes from the non-stationarity of Galactic emissions, hampering the realization of detection methods valid throughout the sky. Still, methods able to efficiently filter out the ``background'' noise have been devised \citep{Leach2008}. The widely used matched filter characterizes the ``background'' noise in terms of its power spectrum. Although this approach has known limitations \citep{HerranzVielva2010} it constitutes a useful reference.

The power spectra at high Galactic latitude ($|b|\ge 30^\circ$) of the astrophysical components present in COrE$+$ maps at selected frequencies (70, 150 and 353 GHz) are illustrated in Fig.~\ref{fig:power_spectrum}. They were computed using the most recent version of the Planck Sky Model \citep{Delabrouille2013} with an apodized Galactic mask (see the caption of Fig.~\ref{fig:power_spectrum}). These figures illustrate the relative importance of the various contributions to the fluctuation field and how they vary with frequency.

\begin{figure*}
\includegraphics[width=0.95\textwidth, angle=0]{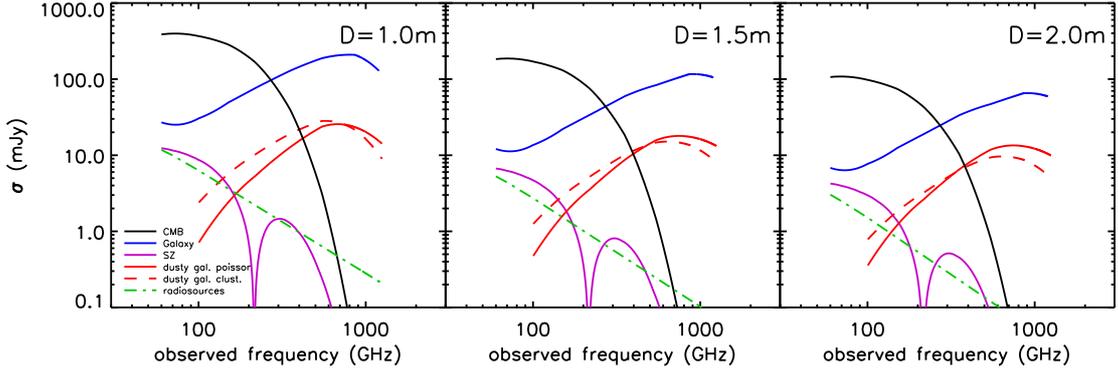}
\caption{Contributions of the various astrophysical components to the root mean square (rms) intensity fluctuations at high Galactic latitude ($|b|\ge 30^\circ$) at the angular resolutions of the three options for the COrE$+$ telescope, as a function of the observed frequency. Fluctuations are dominated by the CMB up to 271 GHz and by Galactic thermal dust at higher frequencies. The dominance of diffuse emissions highlights the importance of efficient component separation algorithms. The WMAP and \textit{Planck} experiences have demonstrated that fluctuations due to diffuse emissions can be largely filtered out by source extraction algorithms.  The contributions of extragalactic sources have been computed after masking sources brighter than the detection limits shown in Fig.~\ref{fig:Slim_vs_sigma}. }
 \label{fig:sigma_vs_freq}
\end{figure*}
%

\begin{figure*}
\includegraphics[width=0.95\textwidth, angle=0]{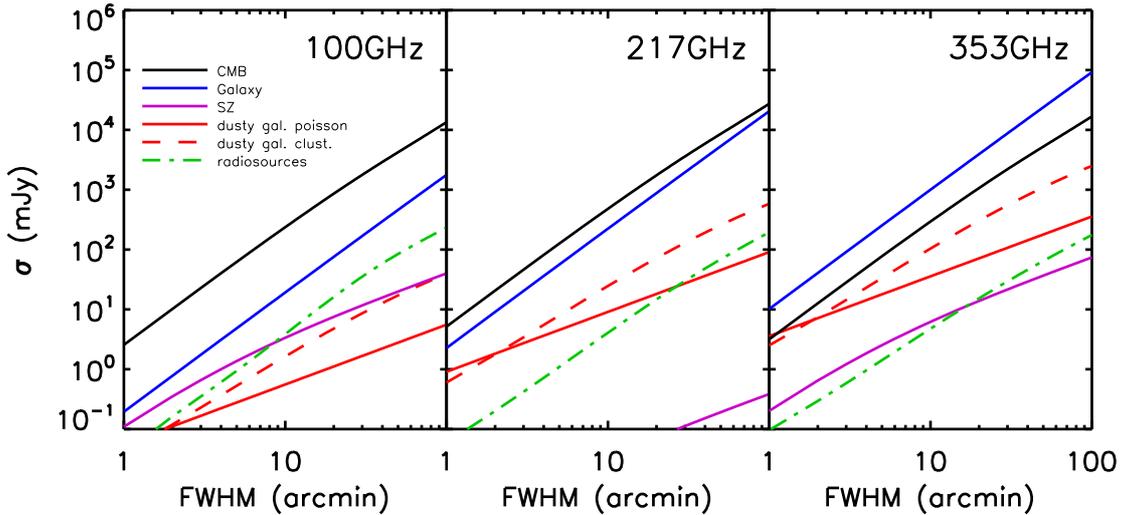}
\caption{Root mean square (rms) intensity fluctuations at the resolution of the measurements as a function of the full width at half maximum (FWHM) of the telescope. The deviation of Poisson fluctuations from the direct proportionality, $\sigma \propto \hbox{FWHM}$, suggested by the combination of eqs.~(\ref{eq:Poisson}) and (\ref{eq:omega_eff}), is due to the decrease of the detection limit, $S_d$, in eq.~(\ref{eq:Poisson}) as the angular resolution improves. The decrease of the fluctuation amplitude with decreasing $S_d$ is stronger for flatter slopes of the source counts.}
 \label{fig:sigma_vs_FWHM}
\end{figure*}

The root mean square (rms) fluctuations, $\sigma$, are related to the power spectrum, $C_\ell$, by \citep{Tegmark1997}
\begin{equation}\label{eq:sigmaq}
\sigma^2=\omega^2 \sum_\ell {2\ell +1\over 4\pi}b_\ell^2 C_\ell \simeq {\omega^2\over 2\pi}{\mathlarger\int}_{\!\!\!0}^\infty d\ell\,\ell\,b_\ell^2 C_\ell ,
\end{equation}
where
\begin{equation}\label{eq:omega}
\omega=2\pi{\mathlarger\int}_{\!\!\!0}^\infty {\rm d}\theta\, \theta\, \exp\left[-{1\over 2}\left({\theta\over \theta_b}\right)^2\right]= 2\pi\theta_b^2
\end{equation}
is the solid angle of the instrument, $b_\ell$ is the experimental beam function, which for a Gaussian beam with standard deviation $\theta_b$ is well approximated by
\begin{equation}
b_\ell=\exp[-{1\over2}\theta_b^2 \ell(\ell +1)].
\end{equation}\label{eq:bell}
The integral approximation to the sum holds if $\theta_b\ll 1\,$rad (so that the relevant $\ell \gg 1$).

Poisson fluctuations are uncorrelated. Hence the only non-zero term of the correlation function is the one at zero-lag and the power spectrum, which is the transform of the correlation function, is independent of $\ell$:
\begin{equation}
C_\ell={\mathlarger\int}_{\!\!\!0}^{S_d} {dN\over dS}\, S^2\, dS ,
\end{equation}\label{eq:flutt}
where $dN(S)/dS$ are the differential number counts per steradian of sources weaker that the detection limit $S_d$. Then
\begin{equation}
{\mathlarger\int}_{\!\!\!0}^\infty d\ell\,\ell\,b_\ell^2 C_\ell = {C_\ell \over 2\theta_b^2}={\pi C_\ell \over \omega}
\end{equation}
and
\begin{equation}\label{eq:Poisson}
\sigma^2\simeq {\omega^2\over 2\pi} {\pi C_\ell \over \omega} = {\omega\over 2} C_\ell =  \omega_{\rm eff} {\mathlarger\int}_{\!\!\!0}^{S_d} {dN\over dS}\, S^2\, dS,
\end{equation}
which is the classical expression for the variance of a Poisson distribution of sources weaker than $S_d$, within a solid angle $\omega_{\rm eff}$ defined by
\begin{equation}
\omega_{\rm eff}=2\pi{\mathlarger\int}_{\!\!\!0}^\infty {\rm d}\theta\, \theta\, f^2(\theta,\phi).
\end{equation}
Adopting a Gaussian instrumental response function (the ``beam'' function), $f(\theta,\phi)$, in polar coordinates:
\begin{equation}\label{eq:beam}
f(\theta,\phi) = \exp\left[-{1\over 2}\left({\theta\over \theta_b}\right)^2\right]\, ,
\end{equation}
we get
\begin{equation}\label{eq:omega_eff}
\omega_{\rm eff}= \pi\theta_b^2={\omega\over 2}={\pi \over 2\, \ln(2)}\left({\hbox{FWHM}\over 2}\right)^2,
\end{equation}
where we have used the relation $\hbox{FWHM}=2\sqrt{2\ln 2}\theta_b\simeq 2.355 \theta_b$. Note that the solid angle to be used to compute Poisson fluctuations is half of that to be used in the general case (eqs.~\ref{eq:sigmaq} and \ref{eq:omega}).

\begin{figure}
\includegraphics[width=0.48\textwidth, angle=0]{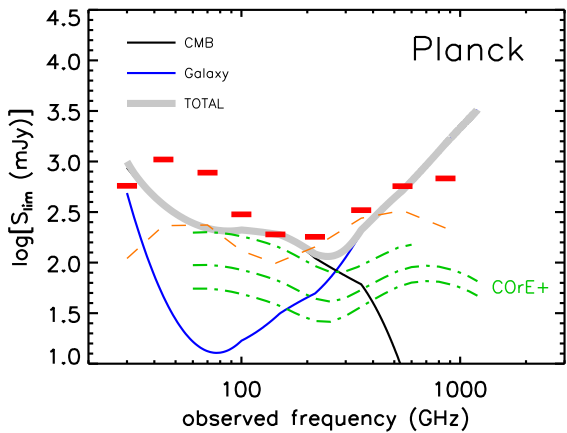}
\begin{minipage}[b]{8.cm}
\caption{PCCS 90\% completeness limits \citep[thick red bars;][]{PCCS} compared with $1\,\sigma$ fluctuations due to CMB and Galactic emissions. The latter have been estimated using the mask described in the caption of Fig.~\ref{fig:power_spectrum}. The fact that the detection limits are generally not far from such rms fluctuations, while the threshold signal to noise ratios for source detection range from 4 to 4.9, illustrates the power of source extraction algorithms in filtering out diffuse emissions. The estimated $4\,\sigma$ detection limits for the 3 COrE$+$ options (dot-dashed green lines) are also shown for comparison. The dashed orange curve shows the $5\sigma$ noise levels for the \textit{Planck} nominal mission (15 months). } \label{fig:Slim_vs_sigma}
\end{minipage}
\end{figure}

The dependence of rms fluctuations on frequency for diffraction limited observations with a 1\,m, 1.5\,m and a 2\,m telescope is illustrated by Fig.~\ref{fig:sigma_vs_freq}. Figure~\ref{fig:sigma_vs_FWHM} shows the trend of rms fluctuations with angular resolution at 3 frequencies. The relative importance of fluctuations due to unresolved point sources increases with frequency and with decreasing FWHM. Their dominant contribution in the FWHM range of interest here comes from radio sources up to about 150 GHz and from dusty galaxies at higher frequencies. While the radio sources have an essentially Poisson distribution, in the case of dusty galaxies the contribution of clustering dominates except on the smallest angular scales.

The above discussion illustrates why, in spite of the very high sensitivity of modern CMB experiments, they only provide shallow surveys of extragalactic sources. For example, the 90\% completeness limits of the \textit{Planck} Catalogue of Compact Sources \citep[PCCS;][]{PCCS} at 70, 100, 217 and 857 GHz are 776, 300, 180 and 680 mJy, respectively,   while the $5\sigma$ instrumental noise limits at the same frequencies, for the 15 month survey, are  $\simeq 235$, 131, 143, and 220 mJy, respectively \citep{Mennella2011,PlanckHFICoreTeam2011}. This means that the surveys are limited by fluctuations of sky signals, not by instrumental noise: there is a lot of useful information in the fluctuation field! A comparison between the PCCS 90\% completeness limits at all \textit{Planck} frequencies and the corresponding $5\sigma$ instrumental noise levels is presented in Fig.~\ref{fig:Slim_vs_sigma}.

A further cue comes from the consideration of the number of detections per resolution element. For example,  the PCCS has 191, 629, 1409 and 6773 detections  at 70, 100, 217 and 857 GHz, respectively, above the 90\% completeness level in the extragalactic zone. This means that there are  1940, 1119, 1868 and 516 resolution elements per source, well above the values (30-40 resolution elements/source) usually corresponding to the confusion limit.

A key issue is the ability of source extraction algorithms to filter out fluctuations due to diffuse emissions, such as the CMB and Galactic radiations.  Figure~\ref{fig:Slim_vs_sigma} shows that the component separation algorithms developed for the \textit{Planck} project  \citep{Leach2008,PlanckCollaborationXII2013} are quite efficient at doing that: over a broad frequency range the PCCS 90\% completeness limits are close to the $1\,\sigma$ fluctuations due to diffuse emissions. In other words most of their contributions to fluctuations are filtered out.

\subsection{Estimate of COrE$+$ detection limits}

To estimate the detection limits for the three COrE$+$ options considered in this paper we have carried out realistic simulations of sky maps including all the relevant foregrounds as well as the instrumental noise for a three year mission. Simulations with resolutions $\ge 4\,$arcmin have been carried out using the latest version of the Planck Sky Model \citep[PSM:][]{Delabrouille2013}.
The PSM includes, in addition to a model of the CMB, Galactic diffuse emissions (synchrotron, free-free, thermal and spinning dust, CO lines), Galactic HII regions, extragalactic radio sources, dusty galaxies, thermal and kinetic Sunyaev-Zeldovich signals from clusters of galaxies. Each component is simulated by means of educated interpolations/extrapolations of the available data, complemented by state-of-the-art models.

Distinctive features of the simulations are spatially varying spectral properties of synchrotron and dust; different spectral parameters for each point source; modelling of the clustering properties of extragalactic sources and of the power spectrum of fluctuations in the cosmic infrared background (CIB), in close agreement with the latest observational determinations.

On angular scales below 4\,arcmin, reached at frequencies above $\simeq 160$--320 GHz (depending on the telescope size), the PSM has not been well tested yet. So, at high frequencies we have performed simpler simulations, taking advantage of the fact that, on small scales, the fluctuation field at high Galactic latitudes is dominated by extragalactic sources making up the CIB and fluctuations due to the CMB and to diffuse Galactic emissions can safely be neglected. The method developed by \citep{GonzalezNuevo2005} was used to distribute the sources on the sky consistent with the measured CIB power spectra at different sub-mm wavelengths \citep{PlanckCollaborationXVIII2011,PlanckCollaborationXXX2013,Viero2013}, i.e. properly allowing for their clustering properties.

The source detection on simulated maps was performed using the IFCAMEX detection pipeline\footnote{http://max.ifca.unican.es/IFCAMEX} and specifically its implementation of the second member of the Mexican Hat Wavelet family (MHW2). This wavelet is obtained applying the Laplacian operator two times on the two-dimensional Gaussian function. The MHW2 is used as a filter to reduce the contribution from the background, including both the small scale noise and the large scale diffuse emission from our own Galaxy, thus enhancing the detection efficiency. The analysis was carried out in projected flat $3.66^\circ \times 3.66^\circ$ square patches, corresponding $128\times 128$ pixels for HEALPix\footnote{http://healpix.sourceforge.net} \citep[Hierarchical Equal Area isoLatitude Pixelation;][]{Gorski2005} maps with $N_{\rm side}=2048$.

For this purpose the map was divided into partially overlapping flat patches. Dividing the sky map into small patches we can optimize the filter taking into account the statistical properties of the map in the vicinity of each source. In practice, we did first a blind run and then a second non-blind run at the position of each source detected in the blind run. The second step allowed us to remove artifacts in the blind catalogue, mostly due to border effects in the filtered images.

Since the flat patches overlap, we can have multiple detections of the same source. The repetitions are removed, keeping for each source only the detection with the highest signal-to-noise ratio. Further details on the MHW2 and the IFCAMEX implementations can be found in \citep{GonzalezNuevo2006} and in \citep{LopezCaniego2006}. The IFCAMEX detection pipeline was previously used to build the New Extragalactic WMAP Point Source (NEWPS) catalogue \citep{LopezCaniego2007,Massardi2009} and the part of the PCCS at $\nu \le 70\,$GHz; it also being used to build the second \textit{Planck} catalogue of compact sources (PCCS2) at the same frequencies. A different implementation of the same algorithm was used for the PCCS and PCSS2 catalogues at higher frequencies.

\begin{figure}
\includegraphics[width=0.45\textwidth, height=6.8cm, angle=90]{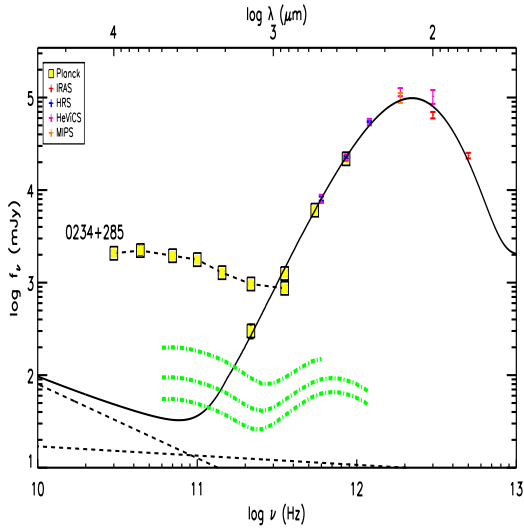}
\begin{minipage}[b]{8.cm}
\caption{Spectral energy distributions (SEDs) of a bright radio source ([HB89]~0234+285; dashed black line) and of a bright infrared galaxy (M61; solid black line) compared with the estimated $4\,\sigma$ detection limits for the three COrE$+$ telescope sizes (dot-dashed green lines). The radio emission of M61 was estimated exploiting the mean relationships between IR and radio emissions derived by \citep{Murphy2011}. The synchrotron and free-free contributions are shown by the dashed black lines at the bottom of the figure. Synchrotron dominates at low frequencies. Note that the frequency at which the radio spectrum of [HB89]~0234+285 crosses the dust emission of M61 is only 0.3 dex higher than the crossing frequency of its own radio emission, which is almost 2 orders of magnitude fainter.}  \label{fig:SED_radio_FIR}
\end{minipage}
\end{figure}

The $4\,\sigma$ detection limits, $S_d$, that, according to our simulations, correspond to $\simgt 90\%$ completeness and reliability in the ``extragalactic zone'' ($|b|>30^\circ$), where Galactic emissions are quite low, are shown in Figs.~\ref{fig:Slim_vs_sigma} and \ref{fig:SED_radio_FIR}. The values of $S_d$ obtained from the simulations are well approximated by the formula:
\begin{equation}\label{eq:Slim}
S_d=4[\sigma_{\rm conf}^2+\sigma_{\rm noise}^2+(0.12\,\sigma_{\rm CMB})^2]^{1/2},
\end{equation}
with
\begin{equation}\label{eq:omega_eff}
\sigma_{\rm conf}^2=\sigma_{\rm P, radio}^2+\sigma_{\rm P, dusty}^2+\sigma_{\rm clust, dusty}^2+\sigma_{\rm SZ}^2.
\end{equation}
In the above formulae the various contributions to the fluctuation field have been computed using the equations given in Sect.~\ref{sec:basics}. The CMB power spectrum corresponds to the \textit{Planck} best fit cosmological parameters \citep{planck_parameters:2013}. The rms confusion fluctuations, $\sigma_{\rm conf}$, include the contributions of radio sources, of dusty galaxies and of galaxy clusters (Sunyaev-Zeldovich, SZ, effect). The Poisson contribution from radio sources was computed via eq.~(\ref{eq:Poisson}) using the \citep{DeZotti2005} model at frequencies of up to 100 GHz and the \citep{Tucci2011} model at higher frequencies. The detection limit, $S_d$, was computed iteratively. A check with the PSM has shown that the clustering of radio sources adds a negligible contribution (we recall that the PSM contains, down to faint flux density levels, real radio sources, at their real positions in the sky). The contributions of dusty galaxies and of galaxy clusters were computed using, respectively, the model by \citep{Cai2013}, that fits accurately the measured CMB power spectra, and the power spectrum of the thermal SZ effect measured by \citep{PlanckCollaborationXXI2014}. These power spectra comprise both the Poisson and the clustering contributions.

%

\begin{figure*}
\includegraphics[width=0.95\textwidth, angle=0]{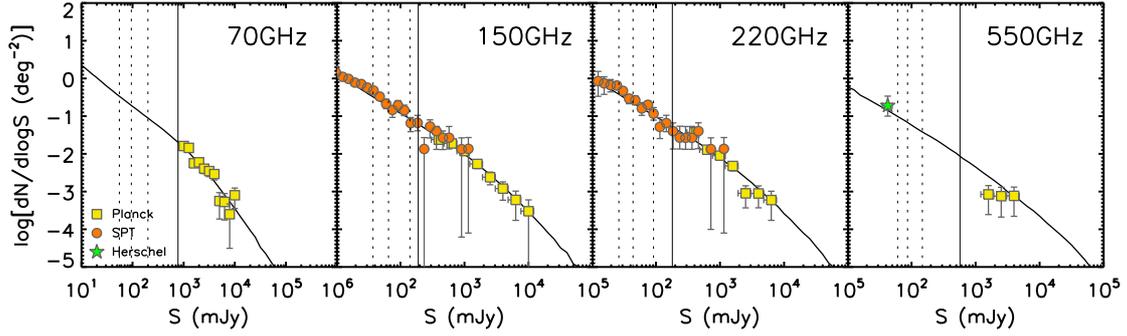}
\caption{Differential number counts of radio sources at 70, 150, 220 and 550 GHz. The \textit{Planck} data points are from \citep{PlanckCollaborationXIII2011} at 70 GHz and from \citep{StatProp2013} at higher frequencies (counts based on the PCCS are not available yet; the 90\% completeness limits of the latter catalogue in the ``extragalactic zone'' are indicated by the vertical solid lines). The South Pole Telescope (SPT) points are from \citep{Mocanu2013} and the \textit{Herschel} point at 550 GHz is from \citep{LopezCaniego2013}. The curves are the model predictions by \citep{DeZotti2005} at 70 GHz and by \citep{Tucci2011} at higher frequencies. The vertical dashed lines correspond to the $4\,\sigma$ detection limits for the 3 COrE$+$ telescope options. }
 \label{fig:radio_counts}
\end{figure*}
%

\begin{figure*}
\includegraphics[width=0.95\textwidth, angle=0]{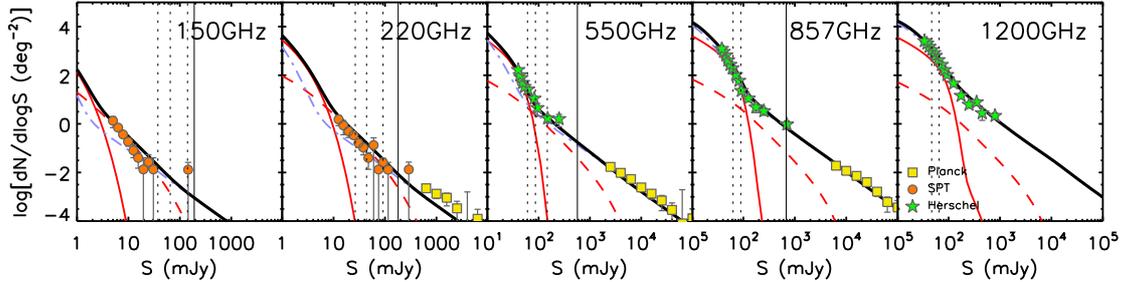}
\caption{Differential number counts of dusty galaxies at 150, 220, 550, 850 and 1200 GHz (2\,mm, 1.4 mm, $545\,\mu$m, $353\,\mu$m and $250\,\mu$m). The \textit{Planck} ERCSC data points are from \citep{StatProp2013} and \citep{Negrello2013}. The SPT points are from \citep{Mocanu2013} and the \textit{Herschel} points from \citep{Clements2010}. The curves are the model predictions by \citep{Cai2013}: the red lines show the contributions of unlensed (solid) and strongly lensed (dashed) proto-spheroidal galaxies, the cyan lines show the counts of late-type normal and starburst galaxies, the back lines are the total.  The vertical dashed lines correspond to the $4\,\sigma$ detection limits for the 3 COrE$+$ telescope options while the vertical solid lines indicate the 90\% completeness limits of the PCCS in the ``extragalactic zone''. }
 \label{fig:IR_counts}
\end{figure*}

\section{Extragalactic sources in the COrE$+$ frequency range}\label{sect:general}

By a lucky (for CMB studies) coincidence the CMB peak occurs at a frequency close to a deep minimum of the intensity of the extragalactic background light as well as of the Galactic emissions. The steep increase with frequency of the dust emission spectrum in the mm/sub-mm region (typically $S_\nu \propto \nu^{3.5}$) while the radio emissions (synchrotron and free-free) decline makes the crossover frequency between radio and dust emission components only weakly dependent on their relative intensities. Moreover, dust temperatures tend to be higher for distant high luminosity sources, partially compensating for the effect of redshift. As a consequence there is an abrupt change in the populations of bright sources above and below $\sim 1\,$mm: radio sources dominate at longer wavelengths, while in the sub-mm region dusty galaxies take over (Fig.~\ref{fig:SED_radio_FIR}).

The \textit{Planck} Early Release Compact Source Catalogue \citep[ERCSC;][]{ERCSC} reported sources extracted from maps built using the data obtained from the scans of the sky between 2009 August 13 and 2010 June 6, comprising the first all-sky survey and about 60\% of the second. It was exploited by \citep{PlanckCollaborationXIII2011} to derive counts of extragalactic radio sources in the range 30--217 GHz and by \citep{StatProp2013} to investigate the statistical properties of infrared and radio extragalactic sources between 100 and 857 GHz.

Subsequent releases of the \textit{Planck} compact source catalogue benefited of additional data (15 months, ``the nominal mission'', for the PCCS, full mission for the PCCS2) as well as of improved data processing. As a result, the completeness limits of the PCCS are fainter than those of the ERCSC by average factors of $\simeq 1.5$ at LFI frequencies (30--70\,GHz) and of $\simeq 2.5$ in the range 100--353\,GHz; the improvement is larger at higher frequencies: a factor of 3.5 at 545 GHz and of 4.5 at 857\,GHz. No information is available yet on the PCCS2. However major improvements are not expected.

Differential counts at several frequencies of radio sources and of dusty galaxies are shown in Figs.~\ref{fig:radio_counts} and \ref{fig:IR_counts}, respectively. The \textit{Planck} counts (yellow data points) were obtained from the ERCSC (the PCCS has not been exploited yet for this purpose). The 90\% completeness limits of the PCCS and of the 3 COrE$+$ options are also shown.  The scientific advances made possible by COrE$+$ are discussed in the following sub-sections.

\subsection{Classical radio sources}

COrE$+$ covers the frequency range where the information on the spectral energy distribution (SED) of radio-loud Active Galactic Nuclei (AGNs) is still scanty: even the SEDs of very bright sources like [HB89]~0234+285 (Fig.~\ref{fig:SED_radio_FIR}) are not measured at sub-mm wavelengths. Yet, in this spectral region important spectral features, carrying essential information on physical conditions of sources, show up.

Observations at mm/sub-mm wavelengths often reveal the transition from optically thick to optically thin radio emission in the most compact regions, i.e. on the maximum self-absorption frequency which roughly corresponds to the synchrotron peak frequency (in terms of $\nu L_\nu$). A systematic survey in the COrE$+$ range will, for example, allow us to see if there are systematic differences in the synchrotron turnover frequencies between BL Lacs and flat-spectrum radio quasars, as would be expected if their jets have  different distributions of the angular separations from the line of sight, implying different amounts of Doppler boosting.  Correlations between turnover frequency and luminosity, which is also boosted by relativistic beaming effects, would help confirm current models.

Major high radio frequency flares have been observed in several compact radio sources \citep{PlanckCollaborationXIV2011,PlanckCollaborationXV2011,Chen2013}, including the recent giant outburst of 3C\,454.3 \citep{Jorstad2010}. Several mechanisms can yield strong variability of relativistically beamed sources \citep[blazars; e.g.][and references therein]{Chen2013}: shocks travelling along the jet, changes of the bulk Lorentz factor causing variations of the Doppler factor, changes of the viewing angle due, e.g., to precession of binary black hole systems or to helical trajectories of plasma elements or to rotating helical jets. Thus variability studies provide clues into the physical properties of emitting regions. Variability due to geometrical effects is expected to affect weakly if at all the source spectra, while shocks induce strong spectral variations with intensity peaks that generally move downwards in frequency with shock age. The COrE$+$ surveys may catch the rise of the flare at the highest frequencies, missed by ground based observations.

As for extended sources, the spectral break frequency, $\nu_b$, at which the synchrotron spectrum steepens due to electron energy losses, is related to the magnetic field, $B$, and to the synchrotron age, $t_s$ (in Myr), by $\nu_b\simeq 96 (30\mu{\rm G}/B)^3\,t_s^{-2}\,$GHz. The systematic multi-frequency study at the COrE$+$ frequencies will thus provide a statistical estimate of the radio source ages. On the other hand, at high redshifts the dominant energy loss mechanism of relativistic electrons may be inverse Compton scattering off CMB photons because the CMB energy density, which grows as $(1+z)^4$, can exceed the magnetic energy density in radio lobes. As a consequence the synchrotron emission of extended, steep-spectrum, sources is suppressed with increasing redshift down to lower and lower frequencies, while compact sources are almost unaffected \citep{Ghisellini2014}. This has obvious implications for the statistics of extended sources as a function of redshift and of the survey frequency. Detailed predictions depend on poorly known quantities, such as the distribution of magnetic field intensities, of source sizes and of injected relativistic electrons. A comparison of low-frequency ground based surveys with the high frequency surveys by COrE$+$ can provide key constraints on these quantities.

Excess far-IR/sub-mm emission due to dust mostly heated by young stars is often observed from radio galaxies, particularly at high $z$ \citep{MileyDeBreuck2008}. COrE$+$'s broad frequency coverage will allow an extensive study of the spectral energy distribution from radio to sub-mm wavelengths and therefore of the presence of dust in host galaxies and of the relationships between the radio emission and the evolutionary status of the host galaxies.

\begin{figure}
\includegraphics[width=0.38\textwidth, angle=90]{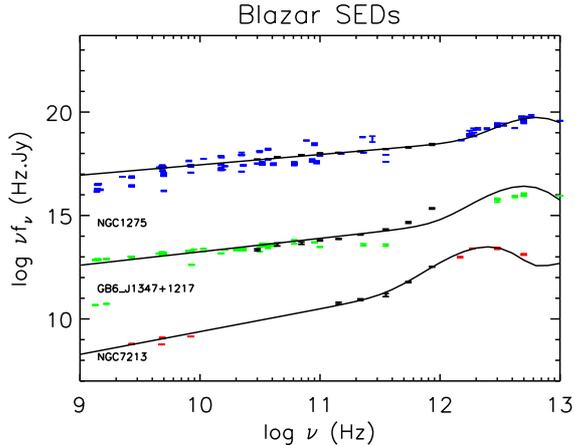}
\begin{minipage}[b]{8.cm}
\caption{SEDs of three  blazars detected by \textit{Planck} showing a sub-mm excess attributable to host galaxies endowed with active star formation. Since, with the exception of galaxies with extreme SFRs, \textit{Planck} could only detect the far-IR emission from nearby galaxies, the sub-mm detection of these blazars implies that a substantial fraction of these sources are hosted by star-forming galaxies, contrary to the notion that there host galaxies are passive ellipticals. The \textit{Planck} data are represented by the black points with error bars; the other data points were taken from the NASA/IPAC Extragalactic Database, NED. For clarity, the SEDs of GB6\,1347$+$1217 and of NGC\,1275 have been raised by $\Delta\log(\nu f_\nu) =3$ and $=6$, respectively. }\label{fig:SED_blazar_dusty}
\end{minipage}
\end{figure}

According to the current notion, host galaxies of blazars are passive ellipticals \citep{Giommi2012}. This view has however already been challenged by \textit{Planck} observations of catalogued blazars associated to galaxies with intense star formation activity (3 examples are shown in Fig.~\ref{fig:SED_blazar_dusty}). Since dust emission from the host galaxy is detectable by \textit{Planck} only for nearby galaxies (but blazars are rare locally) or for extreme IR luminosities, these objects are likely only the tip of the iceberg. The higher sensitivity of COrE$+$ will allow us to extend the study to much larger samples, shedding light on the possible relation of the blazar phenomenon with  star-formation activity.

\subsection{Special radio source populations}

High radio frequency surveys are crucial to investigate special classes of radio sources, self-absorbed up to cm wavelengths. A particularly important class of sources in this category are extreme GHz Peaked Spectrum (GPS) sources or high frequency peakers \citep[HFPs;][]{Dallacasa2000}. GPS sources are powerful [$\log(L_{1.4\rm GHz}/\hbox{erg}\, \hbox{s}^{-1}\, \hbox{Hz}^{-1})\gsim 32]$, compact (sizes $\lsim 1\,$kpc) radio sources with convex spectra peaking at GHz frequencies. Conclusive evidence that they  correspond to the earliest stages of the evolution of powerful radio sources, when the radio-emitting region grows and expands within the interstellar medium of the host galaxy came from VLBI measurements of propagation velocities of up to $\simeq 0.4c$, implying dynamical ages of $\sim 10^3$ years \citep[see][for a review and references]{DeZotti2010}. The identification of these sources is, therefore, a key element in the study of the birth of radio activity.

Predictions for the counts of GPS sources have been worked out by \citep{DeZotti2005,DeZotti2000} and \citep{TintiDeZotti2006}. Observational estimates are complicated by the contamination of candidate GPS samples by flaring blazars that also exhibit spectral peaks at GHz frequencies \citep{Tinti2005}. Recent estimates differ by substantial factors, but the uncertainties are very large due to the poor statistics. \citep{Bonaldi2013} found that only 3 sources with $S_{20\rm GHz} = 200\,$mJy selected over an area of about 6.1\,sr could be classified as genuine GPSs, with a possible incompleteness by a factor of 2. On the other hand, \citep{Hancock2010} have 2 confirmed GPSs with $S_{20\rm GHz} > 200\,$mJy, both with $\nu_p > 80\,$GHz, over a much smaller area; this would correspond to 24--155 (68\% confidence limits for a Poisson distribution) GPSs over 6.1\,sr. Only all sky surveys can produce significant samples of these rare objects, but they need to go deeper than \textit{Planck}. COrE$+$, especially with the largest telescope size, has the right properties. 


\subsection{Local dusty galaxies}

The (sub-)mm surveys performed by all-sky CMB experiments are the ideal way to carry out an unbiased census of dusty galaxies in the local neighbourhood, down to volume densities beyond the reach of more sensitive surveys over small areas of the sky, such as those carried out by \textit{Herschel}. The \textit{Planck} ERCSC has offered the first opportunity to accurately determine the luminosity functions in the very local Universe at several (sub-)millimetre wavelengths, unaffected by cosmological evolution \citep{Negrello2013}, and to investigate their properties \citep{Clemens2013}. These studies have not yet been extended to the PCCS.

In Sect.~\ref{sec:det_lim} we have shown that, at sub-mm wavelengths, COrE$+$ will reach, at 500--600\,GHz flux densities a factor of $\simeq 4$, 6.7 and 9.6, for the 1\,m, 1.5\,m  and 2\,m option, respectively, fainter than the PCCS, i.e. to explore, respectively, a volume a factor of 8, 17 and 30 times larger. This means that COrE$+$ can detect at 600\,GHz, from $\simeq 7,900$ (1\,m option) to $\simeq 17,800$ (1.5\,m option) to $\simeq 31,500$ (2\,m option) star forming galaxies out to $z=0.1$ in the ``extragalactic zone'' .

This redshift range is within that covered by the Sloan Digital Sky Survey (SDSS) whose DR10 release extends over 14,555 square degrees \citep{Ahn2014}. Additional redshifts are provided by other, albeit shallower, large-area spectroscopic redshift surveys: the all-sky 2MASS Redshift Survey \citep[2MRS;][]{Huchra2012} and the IRAS PSC Redshift Survey \citep[PSCz;][]{Saunders2000}; the  hemispherical Six-Degree Field Galaxy Survey \citep[6dFGS;][]{Jones2009}, and others, such as the Two-Degree Field Galaxy Survey \citep[2dFGRS;][]{Colless2003}; see, e.g., \citep{LavauxHudson2011} for a compilation of redshift measurements, dubbed 2M$++$.

\begin{figure*}
\includegraphics[width=0.95\textwidth, angle=0]{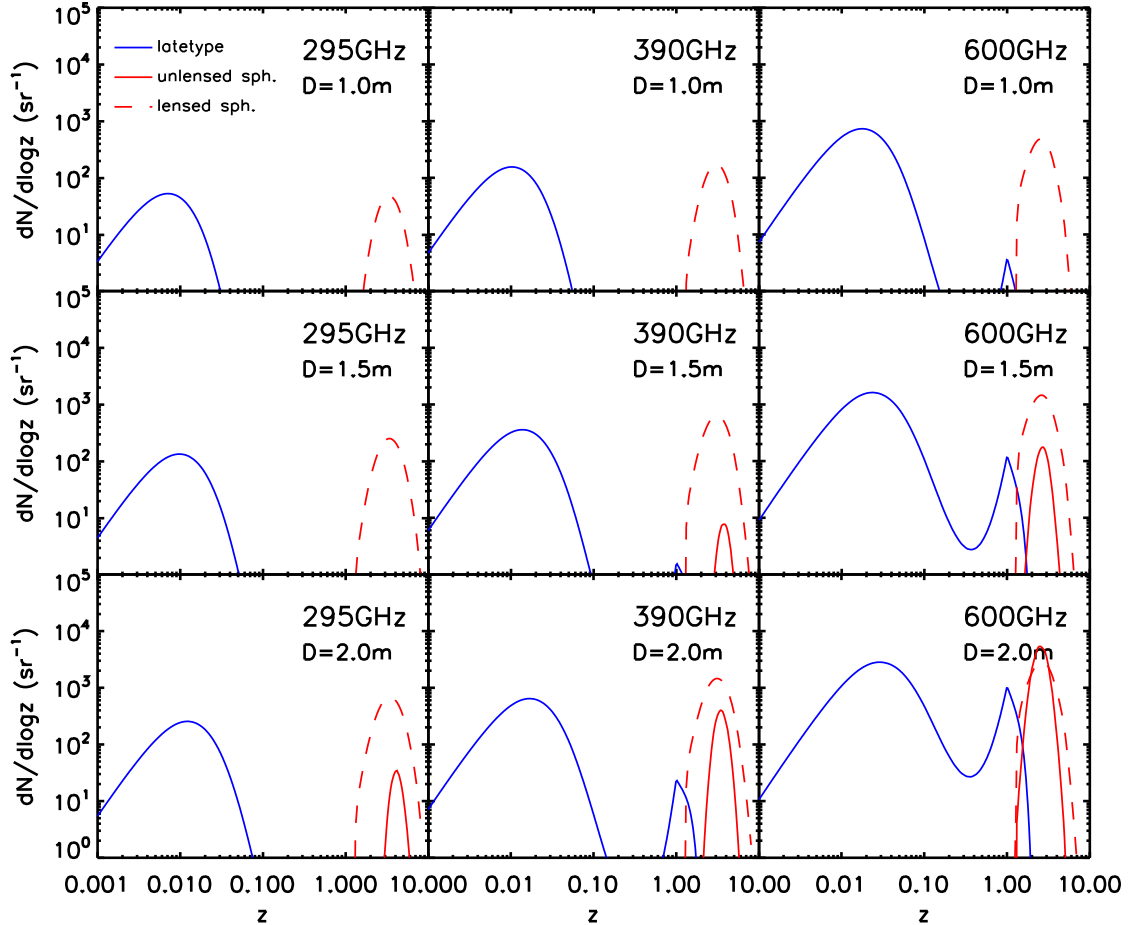}
\caption{Predicted redshift distributions of dusty galaxies at 295, 390 e 600 GHz at the $4\,\sigma$ detection limits of the 1\,m, 1.5\,m and 2\,m options, from top to bottom. Such limits are 81.0, 44.1 and 29.1\,mJy at 295\,GHz; 104.5, 60.7 and 41.6\,mJy at 390\,GHz; 150.5, 90.4 and 63.5\,mJy at 600\,GHz. The distributions are strikingly bi-modal with two well separated peaks. The low-$z$ peak is due to relatively local ($z\lsim 0.1$) star-forming galaxies while the high-$z$ ($z\gsim 1.5$) peak is dominated by strongly lensed galaxies.}
 \label{fig:z_distr}
\end{figure*}

Accurate photometric redshifts have been obtained for many more galaxies detected by wide-angle surveys \citep{Bilicki2014a,Brescia2014,Wang2014}. The 2MASS Photometric Redshift (2MPZ) sample of \citep{Bilicki2014a} should be particularly useful in this respect due to its almost full sky coverage, suitable depth and size (1 million galaxies with a median $z=0.07$; for comparison, the median values of the low-$z$ peaks of the distributions in Fig.~\ref{fig:z_distr} are $\simeq 0.02$--0.03). In addition, COrE$+$ local dusty galaxies missed by 2MASS should be present in the WISE catalogue  \citep{Wright2010};  a large fraction, if not all of them, will have reliable photometric redshifts available over most of the sky \citep[][Bilicki et al. in prep.]{Bilicki2014b}. We expect that with the new data that will be accumulating in the coming years, particularly those from the \textit{Euclid} slitless spectroscopy and from 4-metre Multi-Object Spectroscopic Telescope (4MOST) multifibre $R=5000$ resolution spectroscopy of 10 million $z<1$ emission line galaxies, spectroscopic or photometric redshifts will be available for essentially all the $z<0.1$ galaxies detected by COrE$+$.

The large number of galaxies for which spectroscopic or photometric redshifts will be available will make possible analyses of the properties of galaxies (luminosity functions, dust mass functions, star formation rate functions, dust temperature distributions, etc.) across the Hubble sequence. Of special interest will be the study of IR emission from IR-faint galaxies such as early-type or dwarf galaxies. 
Combining the COrE$+$ with the available multi-wavelength data, especially with those from WISE, 2MASS, GALEX and Euclid we will get a complete view of the SEDs and will be able to investigate relationships between the dust content, the star formation rate (SFR), the stellar mass, the environment and more.

For dusty galaxies, that dominate the extragalactic source population at high frequencies, lower frequency measurements provide information on the relationship, if any, between star-formation and nuclear radio activity and on the radio emission (synchrotron and free-free) powered by star formation.

\subsection{High redshift dusty galaxies}\label{sect:high_z}

Sub-mm surveys \citep[see, e.g.,][for a recent review]{Casey2014} have clearly demonstrated the importance of the FIR/sub-mm data in reconstructing a complete picture of the history of galaxy formation and evolution. The high-redshift sources detected in these surveys are expected to be the progenitors of the present-day massive elliptical galaxies \citep{Lilly1999,Granato2001,Granato2004}. These galaxies can reach extreme IR luminosities \citep[$\ge 10^{13}\,L_\odot$;][]{Lapi2011,Bethermin2011,Gruppioni2013,Barger2014}, that can be further boosted by strong gravitational lensing \citep{Negrello2007,Negrello2010,Negrello2014,Bussmann2013,Messias2014} to the point that some were detected by \textit{Planck}, in spite of the shallowness of its surveys \citep{Fu2012,Combes2012,Herranz2013,Montier2013,Dole2013}.

Using the \citep{Cai2013} model, that accurately fits a broad variety of infrared to millimeter-wave data on extragalactic sources\footnote{See figures in http://staff.ustc.edu.cn/$\sim$zcai/} (multi-frequency and multi-epoch luminosity functions of galaxies and AGNs, redshift distributions, number counts, total and per redshift bins) we find that the deeper COrE$+$ surveys, compared to \textit{Planck}, will detect unlensed IR galaxies up to redshifts $> 1$ (for telescope sizes $\ge 1.5\,$m) and many more strongly lensed (amplification $\mu>2$) galaxies. In the ``extragalactic zone'' ($|b|>30^\circ$) we expect, at 600 GHz, $\simeq 540$ unlensed galaxies at $z>1$ for the 1.5\,m option. For the 2\,m option this number jumps to $\simeq 15,900$.

The counts of strongly lensed galaxies were computed using again the \citep{Cai2013} model, but adopting a maximum amplification $\mu_{\rm max}=30$ to take into account the sources size \citep[at variance with][who assumed point sources]{Cai2013}. As shown by Fig.~2 of \citep{Bonato2014} the adopted $\mu_{\rm max}$ is consistent with the counts and the redshift distribution of strongly lensed galaxies detected by the South Pole Telescope \citep[SPT;][]{Mocanu2013,Weiss2013}.

Even the 1\,m option will detect thousands of strongly lensed galaxies at $|b|>30^\circ$ (about 740 at 390 GHz and over 2,000 at 600 GHz). For the 1.5\,m option these numbers increase to $\simeq 2,900$ and $\simeq 6,200$, respectively; for the 2\,m option they are $\simeq 6,700$ and $\simeq 12,700$, respectively.

For all the options considered the COrE$+$ counts will cover the gap between the bright (sub-)mm counts measured by \textit{Planck} and those measured by the deep SPT and \textit{Herschel} surveys (orange and green points, respectively, in Figs.~\ref{fig:radio_counts} and \ref{fig:IR_counts}). The transition from the bright counts, dominated by nearby galaxies, which have a Euclidean slope, to the fainter ones made much steeper by the combined effect of strong cosmological evolution and of the positive K-correction  \citep{Franceschini1991,BlainLongair1993} provides a strong test for galaxy evolution models. Many of them predict a smoother transition than suggested by present data, but only the much better statistics that will be provided by COrE$+$ will make it possible to reach firm conclusions.


The large samples of strongly gravitationally lensed galaxies that will be provided by COrE$+$ will be of extraordinary astrophysical value in many fields \citep[see][for a review]{Treu2010} thanks to the magnification of the source flux that makes  observable galaxies intrinsically too faint, and to the corresponding increase of the apparent size that makes possible to measure the internal structure of high-$z$ sources to levels otherwise unattainable with the current instrumentation \citep[e.g.,][]{Deane2013}. Follow-up observation will allow us to determine the total (visible and dark) mass of the lensing galaxy, to investigate its density profile, to measure cosmological parameters \citep{Cao2012,Lapi2012,Eales2013,Lubini2014} and especially the Hubble constant using time delays \citep[e.g.,][]{Barnacka2014}.

The selection of strongly lensed galaxies will be extremely easy at about 300--400 GHz. As illustrated by Fig.~\ref{fig:z_distr}, essentially all high-$z$ galaxies detected at these frequencies will be strongly lensed. The other detected sources will be easily recognizable low-$z$ late-type galaxies, plus a small group of radio sources, also easily identifiable in low frequency radio catalogs. At 600\,GHz the fraction of strongly lensed galaxies among $z>1$ objects is still very close to 100\% for the 1\,m option and only slightly decreases  (to 92\%) for 1.5\,m option. For the 2\,m option the strongly lensed fraction among 600\,GHz detections with $z>1$ is 44\%; but in this case we expect $\simeq 6,700$ strongly lensed galaxies in the ``extragalactic zone'' already at 390\,GHz, where they are $\simeq 87\%$ of objects at $z>1$. This means that COrE$+$  will allow us to find very easily thousands of strongly lensed galaxies with a close to 100\% efficiency \citep[see also][]{Negrello2010}.

While other facilities will also be generating large reliable gravitational lens catalogues on a comparable timescale [e.g. Euclid H$\alpha$ lenses \citep{Serjeant2014}, Gaia lensed quasars \citep[e.g.][]{Claeskens2006}], the critical advantages of COrE$+$ will be in extending the sources and lenses to much higher redshifts. This will make it possible to probe the evolution of dark matter halo substructure and the Initial Mass Function (IMF) to much higher redshifts \citep[key tests of semi-analytic models of galaxy evolution, e.g.][]{Vegetti2012,Dutton2013}. This large, high-redshift lens catalogue will also be ideal to probe cosmological parameters using rare double lenses [see, e.g., \citep{CollettAuger2014} for the use of dynamical constraints to remove model degeneracies, e.g. \citep{Schneider2014}]. The background source population will mostly be ultra-luminous star-forming systems that in themselves have always posed the strongest challenges to semi-analytic galaxy evolution models. Only in large lens samples can the rarest high-magnification events be found, and the angular magnification afforded by gravitational lensing in these systems will make them the ideal laboratories for determining the physical processes that dominate star formation and feedback in the early Universe.

\begin{figure}
\includegraphics[width=0.7\textwidth, angle=0]{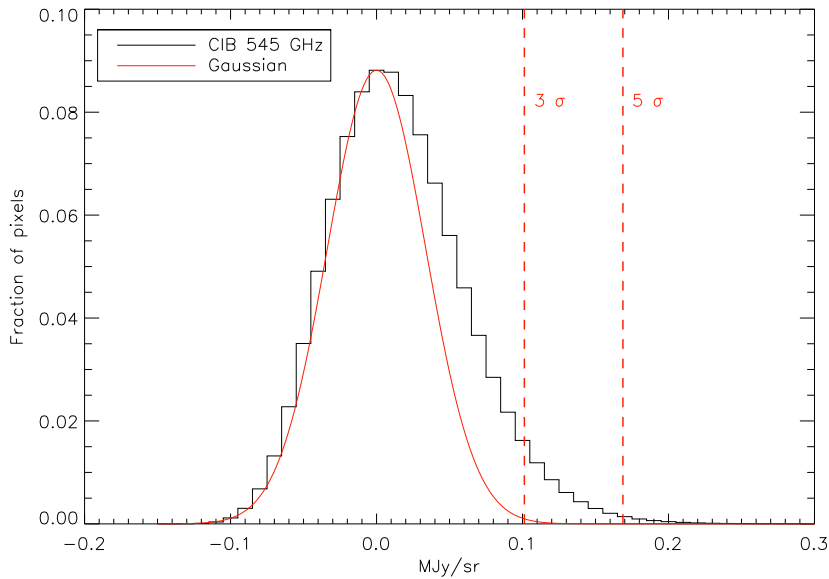}
\begin{minipage}[b]{4.5cm}
\caption{Distribution of intensity peaks of the Cosmic Infrared Background (CIB) at 545 GHz at the \textit{Planck} resolution ($5'$), computed using the PSM. A Gaussian fitting its negative side is also shown for comparison.
\vspace{4.2cm}} \label{fig:CIB_histo}
\end{minipage}
\end{figure}

\section{Extracting information on extragalactic sources below the detection limit }\label{sect:clumps}



\subsection{Proto-clusters of dusty galaxies}
As mentioned in Sect.~\ref{sec:det_lim}, thanks to the very low instrumental noise, the fluctuation field measured by COrE$+$ is signal dominated. This implies that it contains a lot of useful information. In regions with low dust content, several intensity peaks not associated to individual sources just below the detection limit were found to correspond to clumps of dusty galaxies likely to evolve into rich galaxy clusters \citep{Montier2013,Dole2013,Clements2014}. This was predicted by \citep{Negrello2005} on the basis of the argument summarized and slightly updated below.

Sub-mm surveys proved to be most efficient in detecting high-$z$ massive galaxies, interpreted as progenitors of present day giant ellipticals, caught during their  star-formation phase. There is evidence of strong clustering of these sources \citep{Maddox2010,Magliocchetti2011,Magliocchetti2014,Viero2013,PlanckCollaborationXVIII2011,PlanckCollaborationXXX2013}, consistent with them being tracers of strongly overdense regions. Sub-mm surveys are therefore optimally suited to look for proto-clusters at earlier redshifts than can be reached by optical/near-IR, X-ray, SZ surveys.

The analysis in Sect.~\ref{sec:det_lim} has demonstrated that, in high Galactic latitude regions, intensity peaks below the point source detection limits at sub-mm wavelengths are dominated by fluctuations in the distribution of faint sources making up the CIB. Such fluctuation field is highly skewed. One example, computed using the PSM at the \textit{Planck} resolution, is displayed  in Fig.~\ref{fig:CIB_histo}. The clustering contribution overcomes Poisson fluctuations on scales larger than several arcmin (see Fig.~\ref{fig:sigma_vs_FWHM}). This means that sub-mm maps filtered with resolutions of several arcmin are a powerful tool to detect candidate proto-clusters of dusty galaxies.

In other words, the relatively large beams used by CMB experiments collect photons from regions with Mpc physical sizes at $z=1$--3 (at these redshifts the physical scale is $\sim 0.5\,$Mpc/arcmin) thus summing the contributions from tens of star-forming galaxies in over-dense regions. As first pointed out by \citep{Negrello2005} the galaxy clumps may then become detectable, even if individual galaxies are well below the detection limit. Unbiased searches of high-$z$ protoclusters thus become possible, overcoming the need of targeting possible  signposts of high density peaks (high-$z$ radiogalaxies or powerful QSOs), as done so far.

It must be stressed, however, that over-densities are not necessarily proto-clusters. Some may be random alignments, projected on the plane of the sky, of large scale structures either unrelated or connected by a cosmic filament almost aligned with the line of sight. Quantitative predictions were worked out by \citep{Negrello2005}. Briefly, within the standard gravitational clustering scenario, we expect that the distribution of intensity peaks due to source over-densities has a very large variance, resulting from three contributions, discussed below \citep[see][]{Peebles1980}. The mean number of objects inside a volume $V$ centered on a source is:
\begin{equation}
\langle N\rangle_p = n{\mathlarger\int_V}\left[1 + \xi(r)\right]\,dV,
\end{equation}
where $n$ is the mean source number density. The variance around $\langle N\rangle_p$ is:
\begin{eqnarray}
\lefteqn{\langle\left(N  - \langle N\rangle_p\right)^2\rangle_p = \langle N\rangle_p +} \nonumber \\
& & n^2\,{\mathlarger\int_V}{\mathlarger\int_V}\left[\zeta(r_1,r_2) + \xi(r_{12}) - \xi(r_1)\xi(r_2)\right]\,dV_1\,dV_2
\end{eqnarray}
with
\begin{equation}
\zeta(r_1,r_2)=Q(z)\left[\xi(r_1)\xi(r_2)+\xi(r_1)\xi(r_{21})+\xi(r_{12})\xi(r_2) \right],
\end{equation}
$Q(z)$ being the amplitude of the three-point angular correlation function.

For a survey with a Gaussian angular response function [eq.~(\ref{eq:beam})]
%
%
and $\xi(r)=(r_0/r)^{1.8}$ we have
\begin{equation}
nJ_2 = n{\mathlarger\int_V} \xi(r)\,f(\theta)\, dV \simeq 25.9 n\,r_0^{1.8}\,\left[D_A(z)\theta_b\right]^{1.2}
\end{equation}
where $D_A$ is the angular diameter distance and
\begin{equation}
\xi(r,z)=b^2(M_{\rm eff},z)\,\xi_{\rm DM}(r,z)
\end{equation}
wher $b(M_{\rm eff},z)$ is the bias factor, $M_{\rm eff}$ is the effective halo mass, $\,\xi_{\rm DM}(r,z)$ is the spatial correlation function of dark matter halos.
The mean luminosity of the ``clump'' around the central source with luminosity $L_{m}$ is
\begin{eqnarray}
\lefteqn{\bar{L}_{\rm cl}(z)=L_{m}(z) +} \nonumber \\
& & {\mathlarger\int_L}dL'\,L'\,\Phi(L',z)\,\, {\mathlarger\int_V}\left[1 + \xi(r,z)\right]\,f(\theta)\,dV 
\end{eqnarray}
%
with variance:
\begin{eqnarray}
{\sigma^2_{L_{\rm cl}}} =  \sigma^2_L+{\mathlarger\int_L}dL'\,{L'}^2\,\Phi(L',z)\,\, {\mathlarger\int_V}\left[1 + \xi(r,z)\right]\,f(\theta)\,dV  \nonumber \\
 + \left[{\mathlarger\int} dL'\,L'\,\Phi(L',z)\right]^2 \cdot ~~~~~~~~~~~~~~~~~~~~~~~~~~~~~~~~~~~~~~~ \nonumber \\  \!\!\!\!\!\!\!\! {\mathlarger\int}\!\!{\mathlarger\int}\!\!\left[\zeta(r_1,r_2)\! +\! \xi(r_{12})\! -\! \xi(r_1)\xi(r_2)\right]\,f(\theta_1)\,f(\theta_2)\,dV_1\,dV_2,
\end{eqnarray}
where $f(\theta)$ is the response function of the instrument
\begin{eqnarray}
f(\theta) = e^{-(\theta/\Theta)^{2}/2}
\end{eqnarray}
with
\begin{eqnarray}
\Theta = \frac{\rm FWHM}{2\sqrt{2\ln{2}}},
\end{eqnarray}
FWHM being the Full Width at Half Maximum of the instrument. \\
The first contribution to the variance refers to the luminosity of the most luminous source, acting as a  beacon signalling the presence of the proto-cluster; the second refers to the sum of luminosities of the surrounding sources, that sample in a different way the luminosity function; the third refers to the overdensity of neighbouring sources.

All--sky surveys, such as the one planned by COrE$+$, are optimally suited to pick up the rare cases in which all contributions conspire to yield an exceptionally high luminosity of the clump. Particularly favourable situations may occur when the main source is strongly lensed or the clump is within a filament almost aligned with the line of sight, so that the contribution of neighbouring sources (seen in projection) to the observed signal can be very large.

Under the assumption that the statistics of the matter density distribution can be described by a log-normal function \citep{Negrello2005}, the probability distribution function of $L_{\rm cl}$ is:
\begin{equation}
p(L_{\rm cl}, z)={\exp\left[-{1\over 2}\left[\ln(L_{\rm cl})-\mu_g(z)\right]^2/\sigma_g^2(z)\right]\over \sqrt{2\pi\sigma^2_g} L_{\rm cl} }
\end{equation}
where
\begin{eqnarray}
\mu_g(z)&=&\ln\left[{\bar{L}^2_{\rm cl}(z)\over \sqrt{\sigma^2_{L_{\rm cl}}(z) + \bar{L}^2_{\rm cl}(z)} }\right], \nonumber \\
\sigma^{2}_g(z)&=&\ln\left[{\sigma^2_{L_{\rm cl}}(z)\over \bar{L}^2_{\rm cl}(z)}+1\right]\nonumber.
\end{eqnarray}
The clump luminosity function $\Psi(L_{\rm cl},z)$ obeys the normalization condition:
\begin{equation}\mathlarger\int \Psi(L_{\rm cl},z)\, L_{\rm cl}\, dL_{\rm cl}=\mathlarger\int \Phi(L,z)\, L\, dL\ .
\end{equation}
A search for such clumps has been carried out by \citep{Clements2014} by investigating the nature of \textit{Planck} ERCSC sources that lie within an area of $\sim 90\,\hbox{deg}^2$ observed as part of the \textit{Herschel} Multitiered Extragalactic Survey (HerMES). Four sources were found to be associated with overdensities of \textit{Herschel} sources with sub-mm colours suggesting  redshifts $\simeq 1$--2. The estimated surface density of these candidate high-$z$ clumps was found to be consistent with the predictions by \citep{Negrello2005}, although with large uncertainties.

However new calculations taking into account the more recent determinations of the redshift-dependent sub-millimeter luminosity functions (Negrello et al., in preparation) yield lower clump surface densities than estimated by \citep{Negrello2005}, and fall short of accounting for the \citep{Clements2014} result. A very similar conclusion was reached by \citep{Granato2014} using a completely independent approach, coupling hydro-dynamical zoom-in simulations with the recently developed radiative transfer code
GRASIL3D. The origin of the discrepancy is unclear. On one side it may suggest that models and simulations underestimate the IR luminosities of the clumps. Alternatively, the overdensities discovered by  \citep{Clements2014} might be not individual proto-clusters but positive fluctuations in the number of proto-clusters within the \textit{Planck} beam. In both cases the study of these overdensities provides important clues on the early evolution of large scale structure.


\subsection{Power spectrum of the Cosmic Infrared Background (CIB)}


The better angular resolution, at high frequencies, of all COrE$+$ options, compared to \textit{Planck}, will allow us to measure, in a uniform way, the CIB power spectrum over an unprecedented range of frequencies and of angular scales (from $\sim 1\,$arcmin to tens of degrees), thus breaking the degeneracy between the Poisson contribution and that of non-linear effects (one-halo term), that complicates the interpretation of \textit{Planck} measurements, without resorting to external (\textit{Herschel}) data. Although there is good agreement between the latest determinations of the power spectra by \textit{Planck} and \textit{Herschel} \citep{PlanckCollaborationXXX2013,Viero2013}, at 545 GHz the \textit{Herschel} data tend to be slightly but systematically higher, although compatible within the error bars.

Accurate determinations of the CIB power spectrum at different frequencies provide on one side constraints on the evolution of the cosmic star formation density and, on the other side, on halo masses associated to sources of the CIB. Except on the smallest angular scales, the CIB power spectrum is determined by galaxy clustering. The currently standard approach to describe it starts from the consideration that galaxies are hosted by dark matter halos. The power spectrum of the galaxy distribution is parameterized as the sum of the one-halo term that dominates on small scales and depends on the distribution of galaxies within the same halo, and the two-halo term that dominates on large scales and is related to correlations among different halos.

The halo occupation distribution, which is a statistical description of how dark matter halos are populated with galaxies, is modelled using a central-satellite formalism \citep[see, e.g.,][]{Zheng2005}. The two-halo term is essentially proportional to the square of the effective bias factor times the linear theory power spectrum of dark matter halos. Since the bias factor is a steep function of the halo mass and redshift, the two-halo term is a sensitive probe of the effective halo mass of CIB sources (i.e. of galaxies that contribute most to the cosmic star formation) and of its evolution with redshift.

The large number of COrE$+$ channels will make it possible to investigate in detail the decorrelation of power spectra measured in different frequency bands. A decorrelation is expected because the redshift distribution of sources of the CIB shifts to higher and higher redshifts with decreasing frequency \citep[][]{Lapi2011,Xia2012,Cai2013,Bethermin2013}, and was observed \citep{PlanckCollaborationXXX2013,Viero2013}. Accurate measurements of the CIB cross spectra for different frequency channels sets strong constraints on the frequency dependence of redshift distributions, hence on the evolution of the cosmic SFR.

\section{Polarization of extragalactic sources}\label{sect:polarization}

\subsection{Radio sources}

Based on our calculations, we expect 90\% completeness limits in polarized flux density that are substantially lower than the corresponding limits in total intensity, shown in Fig.~\ref{fig:Slim_vs_sigma}, and not far from the $5\,\sigma$ noise levels. This is because, in the COrE$+$ frequency range, the sky is much less complex in polarization than in temperature: the free-free, the spinning dust, the SZ and the CIB emissions are either unpolarized or very weakly polarized; the CMB is also  weakly polarized and, moreover, has a different polarization pattern (dominated by the E-mode) than foreground emissions, which have comparable E- and B-mode polarization. The source confusion level is also low since, roughly, the rms polarization fluctuations, $\sigma_p$, are related to the fluctuations in total intensity, $\sigma_i$, by $\sigma_p\simeq \Pi \sigma_i$, where $\Pi$ is the mean polarization degree \citep{DeZotti1999}. \citep{Massardi2013} found, for flat-spectrum radio sources that dominate the counts at high radio frequencies, $\Pi\simeq 0.026$ at 18\,GHz.

Little is known on the polarization degree of dusty galaxies, but it is likely to be low because the complex structure of galactic magnetic fields with reversals along the line of sight and the disordered alignment of dust grains reduce the global polarized flux when integrated over the whole galaxy. The measurements at $850\,\mu$m of M82 by \citep{GreavesHolland2002} gave a global net polarization degree of only 0.4\%.

The estimated COrE$+$ 90\% completeness limits in polarization for the 1\,m option decrease from $\simeq 25\,$mJy to $\simeq 6\,$mJy as the frequency increases from 60 to 200 GHz and increase at higher frequencies up to $\simeq 23\,$mJy at 350\,GHz. For the 1.5\,m option the $5\,\sigma$ noise levels in polarization vary from  $\simeq 11\,$mJy at 60\,GHz to $\simeq 3\,$mJy at 200\,GHz, to $\simeq 8\,$mJy at 350\,GHz.

Adopting the slope of 1.3 for the integral counts of radio sources in terms of polarized flux density ($N(>S_p)\propto S_p^{-1.3}$) found by \citep{Massardi2013} at 18\,GHz \citep[see also][]{TucciToffolatti2012}, we find that COrE$+$ will detect up to a factor of 40 (1\,m option) or of 160 (1.5\,m option) more polarized radio sources than can be detected by \textit{Planck}. In other words, COrE$+$ will provide blindly selected samples of thousands of polarized radio sources. We caution that these estimates are only tentative because of the complex spectral behaviour of the polarized flux density \citep{Massardi2013} that make extrapolations in frequency quite uncertain. On the other hand, this uncertainty adds interest to COrE$+$ measurements.

\subsection{Dusty galaxies}

If the net polarization degree of M82 is typical of star forming galaxies, we expect the detection in polarization of $\simeq 7.5\,\hbox{sr}^{-1}$ dusty galaxies for the 1\,m option and of $\simeq 37\,\hbox{sr}^{-1}$ for the 1.5\,m option, corresponding to several tens or a few hundred galaxies, respectively, in the ``extragalactic zone'' ($|b|> 30^\circ$). This will provide valuable constraints on the global properties of magnetic fields in star forming galaxies and on their relationships with SFRs.



\section{Discussion and conclusions}\label{sect:conclusions}

COrE$+$ will perform substantially better than \textit{Planck} also in the case of a telescope of a similar or somewhat smaller (1\,m) size. There are two main reasons for this. COrE$+$ will be confusion limited up to the highest frequencies, in contrast to \textit{Planck}. For example, at 545 GHz ($550\,\mu$m) the \textit{Planck} beam has a $\hbox{FWHM} = 3.8'$, while the diffraction limit is $1.5'$. The better resolution implies substantially deeper point source detection limits because the survey depth is mostly limited by fluctuations of sky signals. Realistic simulations give, at 545 GHz, a 90\% completeness level of 96\,mJy for the COrE$+$ 1.5\,m option and of 141\,mJy for the 1\,m option. For comparison the corresponding flux density limit for the PCCS is of 570\,mJy.

At low frequencies, where \textit{Planck} resolution is also at the diffraction limit, the better performance by COrE$+$ is due to its lower instrumental noise. Source detection algorithms efficiently filter out diffuse emissions (CMB and Galaxy), whose power spectra sink down rapidly at the relevant resolutions, but suffer instrumental noise, which has a flat, roughly white noise, power spectrum. Our PSM-based simulations give, at 100 GHz, 90\% completeness limits of 85 and 180\,mJy for the 1.5\,m and 1\,m options, respectively, to be compared with the limit of 300\,mJy for the PCCS.

At higher frequencies, the \textit{Planck} ERCSC has already provided large samples of dusty galaxies mostly at distances $< 100\,$Mpc, thus offering the first opportunity to accurately determine the local luminosity function unaffected by cosmological evolution and to investigate their properties. The PCCS reaches distances about a factor 1.55 larger, increasing the explored volume by about a factor of 3.7. COrE$+$ will reach still further, allowing us to explore a volume a factor of about 17 (1.5\,m option) or of about 8 (1\,m option) larger than the PCCS. Spectroscopic or photometric redshifts will be available for essentially all these galaxies thanks to the large area surveys from the ground and to the \textit{Euclid} slitless spectroscopy. This will make analyses of the properties of galaxies (luminosity functions, dust mass functions, star formation rate functions, dust temperature distributions, etc.) possible across the Hubble sequence.

But the main improvement of COrE$+$ compared to \textit{Planck} observations in total intensity will be in early galaxy evolution. The sub-mm surveys by COrE$+$ will cover the gap between the essentially Euclidean portion measured by \textit{Planck} and the steep one measured by the deep SPT and \textit{Herschel} surveys. The transition between the two portions provides a strong test for galaxy evolution models. Many of them predict it to be smoother than suggested by present data, but only the much better COrE$+$ statistics will make possible to reach firm conclusions.

\textit{Planck} has already found high-redshift galaxies with extreme intrinsic IR luminosities, further boosted by strong gravitational lensing. Hyperluminous IR galaxies are very interesting in themselves since have always posed one of the strongest challenges to semi-analytic galaxy evolution models. Based on our simulations we expect that COrE$+$ will detect, at 600 GHz, from about $321\,\hbox{sr}^{-1}$ (1\,m option) to about $992\,\hbox{sr}^{-1}$ (1.5\,m option) strongly lensed galaxies, i.e. from a few to several thousands such objects in the high Galactic latitude sky. Such large samples are of extraordinary astrophysical value in many fields because the magnification makes galaxies that are intrinsically too faint become observable, while the corresponding increase of the apparent size makes it possible to measure the internal structure to levels otherwise unattainable. Optical spectroscopy of galaxies acting as lenses can be exploited to measure the mass distribution of their dark matter halos as a function of redshift. This will allow a direct test on the evolution of large-scale structure. Samples of thousands of strongly lensed galaxies are essential for many other astrophysical and cosmological applications, e.g. to measure cosmological parameters. Essentially all high-$z$  dusty galaxies will be strongly lensed. The other sources detected at high frequencies will be easily recognizable low-$z$ late-type galaxies, plus a small group of radio sources, also easily identifiable in low frequency radio catalogs. The selection of strongly lensed galaxies will then be extremely easy.

At sub-mm wavelengths fluctuations on scales of several arcmin are dominated by clustering of high-z massive galaxies, interpreted as progenitors of present day giant ellipticals. Sub-mm surveys are therefore optimally suited to look for proto-clusters at earlier redshifts than can be reached by optical/near-IR, X-ray, SZ surveys. \textit{Herschel} observations of intensity peaks with red sub-mm colours, suggestive of high redshifts, in \textit{Planck} high frequency maps have indeed shown that several of them represent the joint emission from multiple starbursts at high $z$, as expected for a young galaxy cluster.

Again, COrE$+$ can do much better than \textit{Planck}. Its high sensitivity sub-mm maps, filtered at several arcmin resolution corresponding to proto-clusters scales at $z = 1$--3, will be a powerful tool to carry out unbiased searches of the long-sought galaxy proto-clusters whose intergalactic gas has not yet reached the virial temperature (i.e. is not detectable in X-rays or via the SZ effect) and whose member galaxies are dust obscured, i.e. optically faint. Only very large area surveys, such as COrE$+$, can provide enough statistics on these rare objects. Higher sensitivity and higher angular resolution follow-up (sub-)mm measurements will assess the nature of candidate proto-clusters. This would help to solve a central problem in cosmology: how did the large scale structure of galaxies form?

COrE$+$ will also measure, in a uniform way, the CIB power spectrum over an unprecedented range of frequencies and of angular scales (from $\simeq 1'$ to tens of degrees), thus breaking the degeneracy between the Poisson contribution and that of non-linear effects (one-halo term), that complicated the interpretation of \textit{Planck} measurements, without resorting to external (e.g. \textit{Herschel} or SPT) data.

Moreover, COrE$+$ will extend the counts of radio sources compared to \textit{Planck}, most notably at high frequencies, where very little is known. Above 217 GHz, the counts will be determined for the first time over a substantial flux density range with good statistics. This will enable the first investigation of the (sub)-mm spectral energy distribution (SED) and of the evolutionary properties of radio sources at (sub-)mm wavelengths. The vast majority of these sources are expected to be blazars, and the accurate determination of their spectra will allow us to understand how physical processes occurring along relativistic jets shape the SED. We will also get numerous samples of `extreme' radio sources allowing us to investigate the rich phenomenology of radio sources at (sub-)mm wavelengths.

A spectacular advance will be made possible by the high COrE$+$ sensitivity in the blind detection of extragalactic sources in polarization. We estimate that COrE$+$ will detect up to a factor of 40 (1\,m option) or of 160 (1.5\,m option) more polarized radio sources than can be detected by \textit{Planck}. We also expect, for the first time, blind samples of several tens (1\,m option) to a few hundred (1.5\,m option) of star forming galaxies detected in polarization in the ``extragalactic zone''.  These samples will open a new window for studies of the global properties of magnetic fields in star forming galaxies and of their relationships with SFRs.

\acknowledgments
Some of the results in this paper have been derived using the HEALPix \citep{Gorski2005} package. Work supported in part by ASI/INAF agreement n. 2014-024-R.0. This research has made use of the NASA/IPAC Extragalactic Database (NED) which is operated by the Jet Propulsion Laboratory, California Institute of Technology, under contract with the National Aeronautics and Space Administration and of data products from the Wide-field Infrared Survey Explorer, which is a joint project of the University of California, Los Angeles, and the Jet Propulsion Laboratory/California Institute of Technology, funded by the National Aeronautics and Space Administration. MB acknowledges the financial assistance of the South African National Research Foundation (NRF) and of the Polish National Science Centre under contract \#UMO-2012/07/D/ST9/02785.

{ }

\begin{thebibliography}{99}


\bibitem{planck_parameters:2013} Planck Collaboration XVI, P.~A.~R. Ade, N. Aghanim, et al., \emph{Planck 2013 results. XVI. Cosmological parameters}, \emph{A\&A} {\bf 571} (2014) A16.


\bibitem{Delabrouille2013} J. Delabrouille, M. Betoule, J.-B. Melin, et al., \emph{The pre-launch Planck Sky Model: a model of sky emission at submillimetre to centimetre wavelengths}, \emph{A\&A} {\bf 553} (2013) A96.


\bibitem{HerranzVielva2010} D. Herranz and P. Vielva, \emph{Cosmic microwave background images}, \emph{IEEE Signal Processing Magazine} {\bf 27} (2010) 67.

\bibitem{Hogbom1974} J.~A. H{\"o}gbom, \emph{Aperture Synthesis with a Non-Regular Distribution of Interferometer Baselines}, \emph{A\&AS} {\bf 15} (1974) 417.

\bibitem{Stetson1992} P.~B. Stetson,  \emph{Astronomical Data Analysis Software and Systems I} {\bf 25} (1992) 297.

\bibitem{BertinArnouts1996} E. Bertin and S. Arnouts, \emph{SExtractor: Software for source extraction}, \emph{A\&AS} {\bf 117} (1996) 393.

\bibitem{Leach2008} S.~M. Leach, J.-F. Cardoso, C. Baccigalupi, et al., \emph{Component separation methods for the PLANCK mission}, \emph{A\&A} {\bf 491} (2008) 597.
    
\bibitem{Tegmark1997} M. Tegmark, \emph{CMB mapping experiments: A designer's guide}, \emph{Phys. Rev. D} {\bf 56} (1997)  4514.
    
\bibitem{PCCS} Planck Collaboration XXVIII, P.~A.~R. Ade, N. Aghanim, et al., \emph{Planck 2013 results. XXVIII. The Planck Catalogue of Compact Sources}, \emph{A\&A} {\bf 571} (2014) A28.

\bibitem{Mennella2011} A. Mennella, M. Bersanelli, R.~C. Butler, et al., \emph{Planck early results. III. First assessment of the Low Frequency Instrument in-flight performance}, \emph{A\&A} {\bf 536} (2011) A3.

\bibitem{PlanckHFICoreTeam2011} Planck HFI Core Team, P.~A.~R. Ade, N. Aghanim, et al., \emph{Planck early results. VI. The High Frequency Instrument data processing}, \emph{A\&A} {\bf 536} (2011) A6.
    
\bibitem{PlanckCollaborationXII2013} Planck Collaboration XII, P.~A.~R. Ade, N. Aghanim, et al., \emph{Planck 2013 results. XII. Diffuse component separation}, \emph{A\&A} {\bf 571} (2014) A12.

\bibitem{GonzalezNuevo2005} J. Gonz{\'a}lez-Nuevo, L. Toffolatti and F. Arg{\"u}eso, \emph{Predictions of the Angular Power Spectrum of Clustered Extragalactic Point Sources at Cosmic Microwave Background Frequencies from Flat and All-Sky Two-dimensional Simulations}, \emph{ApJ} {\bf 621} (2005) 1.

\bibitem{PlanckCollaborationXVIII2011} Planck Collaboration XVIII, P.~A.~R. Ade, N. Aghanim, et al., \emph{Planck early results. XVIII. The power spectrum of cosmic infrared background anisotropies}, \emph{A\&A} {\bf 536} (2011) A18.

\bibitem{PlanckCollaborationXXX2013} Planck Collaboration XXX, P.~A.~R. Ade, N. Aghanim, et al., \emph{Planck 2013 results. XXX. Cosmic infrared background measurements and implications for star formation}, \emph{A\&A} {\bf 571} (2014) A30.

\bibitem{Viero2013} M.~P. Viero, L. Wang, M. Zemcov, et al., \emph{HerMES: Cosmic Infrared Background Anisotropies and the Clustering of Dusty Star-forming Galaxies}, \emph{ApJ} {\bf 772} (2013) 77.


\bibitem{Gorski2005} K.~M. G{\'o}rski, E. Hivon, A.~J. Banday, et al., \emph{HEALPix: A Framework for High-Resolution Discretization and Fast Analysis of Data Distributed on the Sphere}, \emph{ApJ} {\bf 622} (2005) 759.


\bibitem{GonzalezNuevo2006} J. Gonz{\'a}lez-Nuevo, F. Arg{\"u}eso, M. L{\'o}pez-Caniego, M., et al., \emph{The Mexican hat wavelet family: application to point-source detection in cosmic microwave background maps}, \emph{MNRAS} {\bf 369} (2006) 1603.

\bibitem{LopezCaniego2006} M. L{\'o}pez-Caniego, D. Herranz, J. Gonz{\'a}lez-Nuevo,  et al., \emph{Comparison of filters for the detection of point sources in Planck simulations}, \emph{MNRAS} {\bf 370} (2006) 2047.

\bibitem{LopezCaniego2007} M. L{\'o}pez-Caniego, J. Gonz{\'a}lez-Nuevo, D. Herranz, et al., \emph{Nonblind Catalog of Extragalactic Point Sources from the Wilkinson Microwave Anisotropy Probe (WMAP) First 3 Year Survey Data}, \emph{ApJS} {\bf 170} (2007) 108.


\bibitem{Massardi2009} M. Massardi, M. L{\'o}pez-Caniego, J. Gonz{\'a}lez-Nuevo, et al., \emph{Blind and non-blind source detection in WMAP 5-yr maps}, \emph{MNRAS} {\bf 392} (2009) 733.

\bibitem{DeZotti2005} G. De Zotti, R. Ricci, D. Mesa,  et al., \emph{Predictions for high-frequency radio surveys of extragalactic sources}, \emph{A\&A} {\bf 431} (2005) 893.


\bibitem{Tucci2011} M. Tucci, L. Toffolatti, G. De Zotti and E. Mart{\'{\i}}nez-Gonz{\'a}lez, E., \emph{High-frequency predictions for number counts and spectral properties of extragalactic radio sources. New evidence of a break at mm wavelengths in spectra of bright blazar sources},  \emph{A\&A} {\bf 533} (2011) A57.

\bibitem{Cai2013} Z.-Y. Cai, A. Lapi, J.-Q. Xia, et al., \emph{A Hybrid Model for the Evolution of Galaxies and Active Galactic Nuclei in the Infrared}, \emph{ApJ} {\bf 768} (2013) 21.
    
\bibitem{PlanckCollaborationXXI2014} Planck Collaboration XXI, P.~A.~R. Ade, N. Aghanim, et al., \emph{Planck 2013 results. XXI. Power spectrum and high-order statistics of the Planck all-sky Compton parameter map}, \emph{A\&A} {\bf 571} (2014) A21.

\bibitem{Murphy2011} E.~J. Murphy, J.~J. Condon, E. Schinnerer, et al., \emph{Calibrating Extinction-free Star Formation Rate Diagnostics with 33 GHz Free-free Emission in NGC 69}, \emph{ApJ} {\bf 737} (2011) 67.
    
\bibitem{PlanckCollaborationXIII2011} Planck Collaboration XIII, P.~A.~R. Ade, N. Aghanim, et al., \emph{Planck early results. XIII. Statistical properties of extragalactic radio sources in the Planck Early Release Compact Source Catalogue}, \emph{A\&A} {\bf 536} (2011) A13.


\bibitem{Mocanu2013} L.~M. Mocanu, T.~M. Crawford, J.~D. Vieira, et al., \emph{Extragalactic Millimeter-wave Point-source Catalog, Number Counts and Statistics from $771\,\hbox{deg}^2$ of the SPT-SZ Survey}, \emph{ApJ} {\bf 779} (2013) 61.

\bibitem{StatProp2013} Planck Collaboration Int. VII, P.~A.~R. Ade, N. Aghanim, et al., \emph{Planck intermediate results. VII. Statistical properties of infrared and radio extragalactic sources from the Planck Early Release Compact Source Catalogue at frequencies between 100 and 857 GHz}, \emph{A\&A} {\bf 550} (2013) A133.

    
\bibitem{LopezCaniego2013} M. L{\'o}pez-Caniego, J. Gonz{\'a}lez-Nuevo, M. Massardi, et al., \emph{Mining the Herschel-Astrophysical Terahertz Large Area Survey: submillimetre-selected blazars in equatorial fields}, \emph{MNRAS} {\bf 430} (2013) 1566.

\bibitem{Negrello2013} M. Negrello, M. Clemens, J. Gonzalez-Nuevo, et al., \emph{The local luminosity function of star-forming galaxies derived from the Planck Early Release Compact Source Catalogue}, \emph{MNRAS} {\bf 429} (2013) 1309.

\bibitem{Clements2010} D.~L. Clements, E. Rigby, S. Maddox, S., et al., \emph{Herschel-ATLAS: Extragalactic number counts from 250 to 500 microns}, \emph{A\&A} {\bf 518} (2010) L8.

\bibitem{ERCSC} Planck Collaboration VII, P.~A.~R. Ade, N. Aghanim, et al., \emph{Planck early results. VII. The Early Release Compact Source Catalogue}, \emph{A\&A} {\bf 536} (2011) A7.

\bibitem{PlanckCollaborationXIV2011} Planck Collaboration XIV, P.~A.~R. Ade, N. Aghanim, et al., \emph{Planck early results. XIV. ERCSC validation and extreme radio sources}, \emph{A\&A} {\bf 536} (2011) A14.


\bibitem{Chen2013} X. Chen, J.-P. Rachen, M. L{\'o}pez-Caniego, M., et al., \emph{Long-term variability of extragalactic radio sources in the Planck Early Release Compact Source Catalogue}, \emph{A\&A} {\bf 553} (2013) A107.
    
\bibitem{PlanckCollaborationXV2011} Planck Collaboration XV, J. Aatrokoski, P.~A.~R. Ade, et al., \emph{Planck early results. XV. Spectral energy distributions and radio continuum spectra of northern extragalactic radio sources}, \emph{A\&A} {\bf 536} (2011) A15.

\bibitem{Jorstad2010} S.~G. Jorstad,  A.~P. Marscher, V.~M. Larionov, et al., \emph{Flaring Behavior of the Quasar 3C 454.3 Across the Electromagnetic Spectrum}, \emph{ApJ} {\bf 715} (2010) 362.

\bibitem{Ghisellini2014} G. Ghisellini, A. Celotti, F. Tavecchio, F. Haardt and T. Sbarrato, \emph{Radio-loud active galactic nuclei at high redshifts and the cosmic microwave background}, \emph{MNRAS} {\bf 438} (2014) 2694.

\bibitem{MileyDeBreuck2008} G. Miley and C. De Breuck, \emph{Distant radio galaxies and their environments}, \emph{A\&A Rev.} {\bf 15} (2008) 67.

\bibitem{Giommi2012} P. Giommi, G. Polenta, A. L{\"a}hteenm{\"a}ki, et al., \emph{Simultaneous Planck, Swift, and Fermi observations of X-ray and gamma-ray selected blazars}, \emph{A\&A} {\bf 541} (2012) A160.

\bibitem{Dallacasa2000} D. Dallacasa, C. Stanghellini, M. Centonza and R. Fanti, \emph{High frequency peakers. I. The bright sample}, \emph{A\&A} {\bf 363} (2000) 887.

\bibitem{DeZotti2010} G. De Zotti, M. Massardi, M. Negrello an J. Wall, \emph{Radio and millimeter continuum surveys and their astrophysical implications}, \emph{A\&A Rev.} {\bf 18} (2010) 1.
    
\bibitem{DeZotti2000} G. De Zotti, G.~L. Granato, L. Silva, D. Maino and L. Danese, \emph{An evolutionary model for GHz peaked spectrum sources. Predictions for high frequency surveys}, \emph{A\&A} {\bf 354} (2000) 467.
    
\bibitem{Tinti2005} S. Tinti, D. Dallacasa, G. De Zotti, A. Celotti and C. Stanghellini, \emph{High Frequency Peakers: Young radio sources or flaring blazars?},  \emph{A\&A} {\bf 432} (2005) 31.

\bibitem{TintiDeZotti2006} S. Tinti and G. De Zotti, \emph{Constraints on evolutionary properties of GHz Peaked Spectrum galaxies},  \emph{A\&A} {\bf 445} (2006) 889.

\bibitem{Bonaldi2013} A. Bonaldi, L. Bonavera, M. Massardi and G. De Zotti, \emph{The Planck-ATCA Co-eval Observations project: the spectrally selected sample}, \emph{MNRAS} {\bf 428} (2013) 1845.

\bibitem{Hancock2010} P.~J. Hancock, E.~M. Sadler, E.~K. Mahony and R. Ricci, \emph{Observations and properties of candidate high-frequency GPS radio sources in the AT20G survey}, \emph{MNRAS} {\bf 408} (2010) 1187.

\bibitem{Clemens2013} M.~S. Clemens, M. Negrello, G. De Zotti, et al., \emph{Dust and star formation properties of a complete sample of local galaxies drawn from the Planck Early Release Compact Source Catalogue}, \emph{MNRAS} {\bf 433} (2013) 695.


\bibitem{Ahn2014} C.~P. Ahn, R. Alexandroff, C. Allende Prieto, et al., \emph{The Tenth Data Release of the Sloan Digital Sky Survey: First Spectroscopic Data from the SDSS-III Apache Point Observatory Galactic Evolution Experiment}, \emph{ApJS} {\bf 211} (2014) 17.


\bibitem{Huchra2012} J.~P. Huchra, L.~M. Macri, K.~L. Masters,  et al., \emph{The 2MASS Redshift Survey---Description and Data Release}, \emph{ApJS} {\bf 199} (2012) 26.
    
\bibitem{Saunders2000} W. Saunders, W.~J. Sutherland, S.~J. Maddox, et al., \emph{The PSCz catalogue},  \emph{MNRAS} {\bf 317} (2000) 55.

\bibitem{Jones2009} D.~H. Jones, M.~A. Read, W. Saunders, et al., \emph{The 6dF Galaxy Survey: final redshift release (DR3) and southern large-scale structures}, \emph{MNRAS} {\bf 399} (2009) 683.

\bibitem{Colless2003} M. Colless, B.~A. Peterson, C. Jackson,  et al., \emph{The 2dF Galaxy Redshift Survey: Final Data Release} (2003) [arXiv:astro-ph/0306581]

\bibitem{LavauxHudson2011} G. Lavaux and M.~J. Hudson, \emph{The 2M$++$ galaxy redshift catalogue}, \emph{MNRAS} {\bf 416} (2011) 2840.

\bibitem{Bilicki2014a} M. Bilicki, T.~H. Jarrett, J.~A. Peacock, M.~E. Cluver and L. Steward, \emph{Two Micron All Sky Survey Photometric Redshift Catalog: A Comprehensive Three-dimensional Census of the Whole Sky}, \emph{ApJS} {\bf 210} (2014) 9.

\bibitem{Brescia2014} M. Brescia, S. Cavuoti, G. Longo and V. De Stefano, \emph{A catalogue of photometric redshifts for the SDSS-DR9 galaxies}, \emph{A\&A} {\bf 568} (2014) A126.

\bibitem{Wang2014} L. Wang, M. Rowan-Robinson, P. Norberg, S. Heinis, J. Han, J., \emph{The Revised IRAS-FSC Redshift Catalogue (RIFSCz)}, \emph{MNRAS} {\bf 442} (2014) 2739.


\bibitem{Wright2010} E.~L. Wright, P.~R.~M. Eisenhardt, A. Mainzer, et al., \emph{The Wide-field Infrared Survey Explorer (WISE): Mission Description and Initial On-orbit Performance},  \emph{AJ} {\bf 140} (2010) 1868.


\bibitem{Bilicki2014b} M. Bilicki, J.~A. Peacock, T.~H. Jarrett, M.~E. Cluver and L. Steward, \emph{Mapping the Cosmic Web with the largest all-sky surveys} (2014) [arXiv:1408.0799].

\bibitem{Laureijs2011} R. Laureijs, J. Amiaux, S. Arduini, et al.,  \emph{Euclid Definition Study Report}, (2011) [arXiv:1110.3193].

\bibitem{deJong2012} R.~S. de Jong, O. Bellido-Tirado, C. Chiappini, et al., \emph{4MOST: 4-metre multi-object spectroscopic telescope}, \emph{SPIE Conf. Ser.} {\bf 8446} (2012), 84460T.


\bibitem{Casey2014} C.~M. Casey, D. Narayanan and A. Cooray, \emph{Dusty star-forming galaxies at high redshift}, \emph{Phys. Rep.} {\bf 541} (2014) 45.
    
\bibitem{Granato2004} G.~L. Granato, G. De Zotti, L. Silva, A. Bressan and L. Danese, \emph{A Physical Model for the Coevolution of QSOs and Their Spheroidal Hosts}, \emph{ApJ} {\bf 600} (2004) 580.

\bibitem{Granato2001} G.~L. Granato, L. Silva, P. Monaco, P. Panuzzo, P. Salucci, G. De Zotti and L. Danese, \emph{Joint formation of QSOs and spheroids: QSOs as clocks of star formation in spheroids},  \emph{MNRAS} {\bf 324} (2001) 757.

\bibitem{Lilly1999} S.~J. Lilly, S.~A. Eales, W.~K.~P. Gear, et al., \emph{The Canada-United Kingdom Deep Submillimeter Survey. II. First Identifications, Redshifts, and Implications for Galaxy Evolution}, \emph{ApJ} {\bf 518} (1999) 641.




\bibitem{Barger2014} A.~J. Barger, L.~L. Cowie, C.~C. Chen, et al., \emph{Is There a Maximum Star Formation
Rate in High-redshift Galaxies?}, \emph{ApJ} {\bf 784} (2014) 9.


\bibitem{Bethermin2011} M. B{\'e}thermin, H. Dole, G. Lagache, D. Le Borgne and A. Penin,  \emph{Modeling the evolution of infrared galaxies: a parametric backward evolution model}, \emph{A\&A} {\bf 529} (2011) A4.

\bibitem{Gruppioni2013} C. Gruppioni, F. Pozzi, G. Rodighiero, et al., \emph{The Herschel PEP/HerMES luminosity function - I. Probing the evolution of PACS selected Galaxies to $z\simeq 4$}, \emph{MNRAS} {\bf 432} (2013) 23.

\bibitem{Lapi2011} A. Lapi, J. Gonz{\'a}lez-Nuevo, L. Fan, et al., \emph{Herschel-ATLAS Galaxy Counts and High-redshift Luminosity Functions: The Formation of Massive Early-type Galaxies}, \emph{ApJ} {\bf 742} (2011) 24.
    
\bibitem{Bussmann2013} R.~S. Bussmann, I. P{\'e}rez-Fournon, S. Amber, S., et al., \emph{Gravitational Lens Models Based on Submillimeter Array Imaging of Herschel-selected Strongly Lensed Sub-millimeter Galaxies at $z > 1.5$}, \emph{ApJ} {\bf 779} (2013) 25.
    
\bibitem{Messias2014} Messias, H., Dye, S., Nagar, N., et al., \emph{Herschel-ATLAS and ALMA. HATLAS J142935.3-002836, a lensed major merger at redshift 1.027}, \emph{A\&A} {\bf 568} (2014) A92.

\bibitem{Negrello2010} M. Negrello, R. Hopwood, G. De Zotti, et al., \emph{The Detection of a Population of Submillimeter-Bright, Strongly Lensed Galaxies}, \emph{Science} {\bf 330} (2010) 800.

\bibitem{Negrello2014} M. Negrello, R. Hopwood, S. Dye, et al., \emph{Herschel-ATLAS: deep HST/WFC3 imaging of strongly lensed submillimetre galaxies}, \emph{MNRAS} {\bf 440} (2014) 1999.

\bibitem{Negrello2007} M. Negrello, F. Perrotta, J. Gonz{\'a}lez-Nuevo, et al., \emph{Astrophysical and cosmological information from large-scale submillimetre surveys of extragalactic sources}, \emph{MNRAS} {\bf 377} (2007) 1557.

\bibitem{Combes2012} F. Combes, M. Rex, T.~D. Rawle, T.~D., et al., \emph{A bright $z = 5.2$ lensed submillimeter galaxy in the field of Abell 773. HLSJ$091828.6+514223$}, \emph{A\&A} {\bf 538} (2012) L4.

\bibitem{Dole2013} H. Dole, e al., \emph{Herschel Unveils Enigmatic Planck Extreme high-­$z$ Source Candidates: Overview}, ESA Symp. ``The Universe Explored by Herschel'' (2013)

\bibitem{Fu2012} H. Fu, E. Jullo, A. Cooray, et al., \emph{A Comprehensive View of a Strongly Lensed Planck-Associated Submillimeter Galaxy}, \emph{ApJ} {\bf 753} (2012) 134.

\bibitem{Herranz2013} D. Herranz, J. Gonz{\'a}lez-Nuevo, D.~L. Clements, et al., \emph{Herschel-ATLAS: Planck sources in the phase 1 fields}, \emph{A\&A} {\bf 549} (2013) A31.
    
\bibitem{Montier2013} L. Montier,  \emph{Planck and Herschel}, in  47th ESLAB Symp. ``The Universe as seen by Planck'', (2013)

\bibitem{Bonato2014} M. Bonato, M. Negrello, Z.-Y. Cai, et al., \emph{Exploring the early dust-obscured phase of galaxy formation with blind mid-/far-infrared spectroscopic surveys}, \emph{MNRAS} {\bf 438} (2014) 2547.
    
\bibitem{Weiss2013} A. Wei{\ss}, C. De Breuck, D.~P. Marrone, et al., \emph{ALMA Redshifts of Millimeter-selected Galaxies from the SPT Survey: The Redshift Distribution of Dusty Star-forming Galaxies},  \emph{ApJ} {\bf 767} (2013) 88.

\bibitem{BlainLongair1993} A.~W. Blain and M.~S. Longair, \emph{Submillimetre Cosmology}, \emph{MNRAS} {\bf 264} (1993) 509.

\bibitem{Franceschini1991} A. Franceschini, L. Toffolatti, P. Mazzei, L. Danese and G. De Zotti, \emph{Galaxy counts and contributions to the background radiation from 1 micron to 1000 microns}, \emph{A\&AS} {\bf 89} (1991) 285.

\bibitem{Treu2010} T. Treu, \emph{Strong Lensing by Galaxies}, \emph{ARA\&A} {\bf 48} (2010) 87.

    
\bibitem{Deane2013} R.~P. Deane, I. Heywood, S. Rawlings and P.~J. Marshall, \emph{The preferentially magnified active nucleus in IRAS F10214+4724 - II. Spatially resolved cold molecular gas}, \emph{MNRAS} {\bf 434} (2013) 23.

\bibitem{Cao2012} S. Cao, Y. Pan, M. Biesiada, W. Godlowski and Z.-H. Zhu, \emph{Constraints on cosmological models from strong gravitational lensing systems}, \emph{JCAP} {\bf 3} (2012) 16.

\bibitem{Eales2013} S. Eales, \emph{Practical cosmology with lenses},  \emph{MNRAS} {\bf 446} (2015) 3224.

\bibitem{Lapi2012} A. Lapi, M. Negrello, J. Gonz{\'a}lez-Nuevo,  et al., \emph{Effective Models for Statistical Studies of Galaxy-scale Gravitational Lensing}, \emph{ApJ} {\bf 755} (2012) 46.
    
\bibitem{Lubini2014} M. Lubini, M. Sereno, J. Coles, P. Jetzer and P. Saha, \emph{Cosmological parameter determination in free-form strong gravitational lens modelling}, \emph{MNRAS} {\bf 437} (2014) 2461.


\bibitem{Barnacka2014} A. Barnacka, M. Geller, I.~P. Dell'Antonio and W. Benbow, \emph{Strongly Lensed Jets, Time
Delays, and the Value of H0}  (2014) [arXiv:1408.5898]

\bibitem{Serjeant2014} S. Serjeant, \emph{Up to 100,000 Reliable Strong Gravitational Lenses in Future Dark Energy Experiments},  \emph{ApJ} {\bf 793} (2014) L10

\bibitem{Claeskens2006} J.~F. Claeskens, A. Smette, L. Vandenbuckle and J. Surdej, \emph{Identification and redshift determination of quasi-stellar objects with medium-band photometry: application to Gaia}, \emph{MNRAS} {\bf 367} (2006) 879.
    
\bibitem{Dutton2013} A.~A. Dutton, A.~V. Macci\`o, J.~T. Mendel and L. Simard, \emph{Universal IMF versus dark halo response in early-type galaxies: breaking the degeneracy with the Fundamental Plane}, \emph{MNRAS} {\bf 432} (2013) 2496.

\bibitem{Vegetti2012} S. Vegetti, D.~J. Lagattuta, J.~P. McKean, M.~W. Auger, C.~D. Fassnacht and L.~V.~E. Koopmans, \emph{Gravitational detection of a low-mass dark satellite galaxy at cosmological distance}, \emph{Nature} {\bf 481} (2012) 341.

\bibitem{CollettAuger2014} T.~E. Collett and M.~W. Auger, \emph{Cosmological constraints from the double source plane lens SDSSJ$0946+1006$}, \emph{MNRAS} {\bf 443} (2014) 969.

\bibitem{Schneider2014} P. Schneider, \emph{Can one determine cosmological parameters from multi-plane strong lens systems?}, \emph{A\&A} {\bf 568} (2014) L2.

\bibitem{Clements2014} D.~L. Clements, F.~G. Braglia, A.~K. Hyde, et al., \emph{Herschel Multitiered Extragalactic Survey: clusters of dusty galaxies uncovered by Herschel and Planck}, \emph{MNRAS} {\bf 439} (2014) 1193.

\bibitem{Negrello2005} M. Negrello, J. Gonz{\'a}lez-Nuevo, M. Magliocchetti, et al., \emph{Effect of clustering on extragalactic source counts with low-resolution instruments}, \emph{MNRAS} {\bf 358} (2005) 869.

\bibitem{Maddox2010} S.~J. Maddox, L. Dunne, E. Rigby, et al., \emph{Herschel-ATLAS: The angular correlation function of submillimetre galaxies at high and low redshift}, \emph{A\&A} {\bf 518} (2010) L11.

\bibitem{Magliocchetti2014} M. Magliocchetti, A. Lapi, M. Negrello, G. De Zotti and L. Danese, \emph{Cosmic dichotomy in the hosts of rapidly star-forming systems at low and high redshifts}, \emph{MNRAS} {\bf 437} (2014) 2263.

\bibitem{Magliocchetti2011} M. Magliocchetti, P. Santini, G. Rodighiero, et al., \emph{The PEP survey: clustering of infrared-selected galaxies and structure formation at $z\sim ˜2$ in GOODS-South}, \emph{MNRAS} {\bf 416} (2011) 1105.

\bibitem{Peebles1980} P.~J.~E. Peebles, \emph{The large-scale structure of the universe}, Princeton University Press (1980).
    
\bibitem{Granato2014} G.~L. Granato, C. Ragone-Figueroa, R. Dominguez-Tenreiro, et al., \emph{The early phases of galaxy clusters formation in IR: coupling hydrodynamical simulations with GRASIL3D}, (2014) [arXiv:1412.6105]

\bibitem{Zheng2005} Z. Zheng, A.~A. Berlind, D.~H. Weinberg,  et al., \emph{Theoretical Models of the Halo Occupation Distribution: Separating Central and Satellite Galaxies}, \emph{ApJ} {\bf 633} (2005) 791.

\bibitem{Bethermin2013} M. B{\'e}thermin, L. Wang, O. Dor{\'e}, et al., \emph{The redshift evolution of the distribution of star formation among dark matter halos as seen in the infrared}, \emph{A\&A} {\bf 557} (2013) A66.

\bibitem{Xia2012} J.-Q. Xia, M. Negrello, A. Lapi, et al., \emph{Clustering of submillimetre galaxies in a self-regulated baryon collapse model}, \emph{MNRAS} {\bf 422} (2012) 1324.

\bibitem{DeZotti1999} G. De Zotti, C. Gruppioni, P. Ciliegi, C. Burigana and L. Danese, \emph{Polarization fluctuations due to extragalactic sources}, \emph{New Astr.} {\bf 4} (1999) 481.

\bibitem{Massardi2013} M. Massardi, S.~G. Burke-Spolaor, T. Murphy, et al., \emph{A polarization survey of bright extragalactic AT20G sources}, \emph{MNRAS} {\bf 436} (2013) 2915.

\bibitem{GreavesHolland2002} J.~S. Greaves and W.~S. Holland,  \emph{Submillimetre polarization of M82 and the Galactic Center: Implications for CMB polarimetry}, in Astrophysical Polarized Backgrounds, AIP Conf. Proc. {\bf 609} (2002) 267.

\bibitem{TucciToffolatti2012} M. Tucci and L. Toffolatti, \emph{The Impact of Polarized Extragalactic Radio Sources on the Detection of CMB Anisotropies in Polarization}, \emph{Advances in Astronomy} (2012) 624987.





























\end{thebibliography}
\end{document}